\documentclass[journal,final]{IEEEtran}
\ifCLASSINFOpdf
\else
\fi
\usepackage{epsfig}
\usepackage{graphicx}
\usepackage{amsmath}
\usepackage{amssymb}
\usepackage{textcomp}
\usepackage{mathrsfs}
\usepackage{booktabs}
\usepackage{diagbox}
\usepackage{xcolor}
\usepackage{multirow}
\usepackage{subfigure}
\usepackage{footnote}
\usepackage{booktabs}
\usepackage{threeparttable}
\usepackage{ragged2e}
\usepackage{textcase}

\usepackage{makecell}
\usepackage{array,color}
\newcommand{\ie}{\emph{i.e.}}

\newcommand{\eg}{\emph{e.g.}}
\usepackage{bm}
\makeatletter
\newcommand*\bigcdot{\mathpalette\bigcdot@{.5}}
\newcommand*\bigcdot@[2]{\mathbin{\vcenter{\hbox{\scalebox{#2}{$\m@th#1\bullet$}}}}}
\makeatother

\usepackage[pagebackref=false,breaklinks=true,letterpaper=true,colorlinks,bookmarks=false,linkcolor=red,citecolor=blue,urlcolor=blue]{hyperref}

\makeatletter
\newif\if@restonecol
\makeatother

\usepackage[linesnumbered,ruled,vlined]{algorithm2e}
\usepackage{algpseudocode}
\usepackage{amsmath}

\begin{document}
	%
	\title{\textcolor{black}{Hidden Path Selection Network for \\ Semantic Segmentation of Remote Sensing Images}}
	%
	%
	%
	
	\author{Kunping~Yang, \and Xin-Yi~Tong,
		\and Gui-Song~Xia,~\IEEEmembership{Senior Member,~IEEE,} \\
		\and Weiming Shen, \and Liangpei Zhang,~\IEEEmembership{Fellow,~IEEE} 
		\IEEEcompsocitemizethanks{
			\IEEEcompsocthanksitem The study of this paper is funded by the National Natural Science Foundation of China (NSFC) under grant contracts No.61871299 and No.41820104006. 
			\IEEEcompsocthanksitem K. Yang, W. Shen, and L. Zhang are with the State Key Lab. of LIESMARS, Wuhan University, Wuhan, 430072, China.  Email: \{{\em kunpingyang, shenwm, zlp62}\}@whu.edu.cn.
			\IEEEcompsocthanksitem Xin-Yi Tong is with the Remote Sensing Technology, German Aerospace
			Center (DLR), 82234 Wessling, Germany, and also with the Technical University
			of Munich, 80333 Munchen, Germany. Email: {\em xinyi.tong}@dlr.de.
			\IEEEcompsocthanksitem G.-S. Xia is with the National Engineering Research Center for Multimedia Software, School of Computer Science, Institute of Artificial Intelligence, and also the State Key Lab. LIESMARS, Wuhan University, Wuhan, 430072, China.  Email: {\em guisong.xia}@whu.edu.cn.
			\IEEEcompsocthanksitem Corresponding author: Weiming Shen (shenwm@whu.edu.cn).
		}
	}

	\maketitle
	
	\begin{abstract}
		\justifying
		Targeting at depicting land covers with pixel-wise semantic categories, semantic segmentation in remote sensing images needs to portray diverse distributions over vast geographical locations, which is difficult to be achieved by the homogeneous pixel-wise forward paths in the architectures of existing deep models.
		Although several algorithms have been designed to select pixel-wise adaptive forward paths for natural image analysis, it still lacks 
		theoretical supports on how to obtain optimal selections.
		In this paper, we provide mathematical analyses in terms of the parameter optimization, which guides us to design a method called {\em Hidden Path Selection Network} (HPS-Net).
		With the help of hidden variables derived from an extra mini-branch, HPS-Net is able to tackle the inherent problem about inaccessible global optimums by adjusting the direct relationships between feature maps and pixel-wise path selections in existing algorithms, which we call hidden path selection. 
		For the better training and evaluation, we further refine and expand the 5-class Gaofen Image Dataset (GID-5) to a new one with 15 land-cover categories, \ie, GID-15.
		The experimental results on both GID-5 and GID-15 demonstrate that the proposed modules can stably improve the performance of different deep structures, which validates the proposed mathematical analyses.
	\end{abstract}
	

	%
	\IEEEpeerreviewmaketitle

	\section{Introduction}\label{sec:introduction}
	%
	%
	%
	%
	\IEEEPARstart{S}{emantic} segmentation in remote sensing images~\cite{2018An, 2019Joint, 2020Land, 0Comprehensively}, which targets at identifying land covers with pixel-wise semantic categories, is of considerable significance among many land use management related tasks~\cite{0High, Urban_physics, Arneth2015Climate}. 
	Due to the complexity of real-world environments, tackling issues such as the diversity of land-cover distributions over different geographical locations is an important topic among existing investigations that are related to remote sensing image interpretation~\cite{AID,Million-AID,Gibbs_MRF_TIP,DOTA,DOTAv2PAMI,Slow_feature_wc1,Slow_feature_wc2}. As a consequence, studying semantic segmentation methods to depict diverse complicated land-cover distributions simultaneously has drawn increasing attentions in the remote sensing community \cite{2020Scale, 2020Exploring, 2021Development}.
	
	
	\begin{figure}[!t]
		\centering
		{\includegraphics[width=1\linewidth]{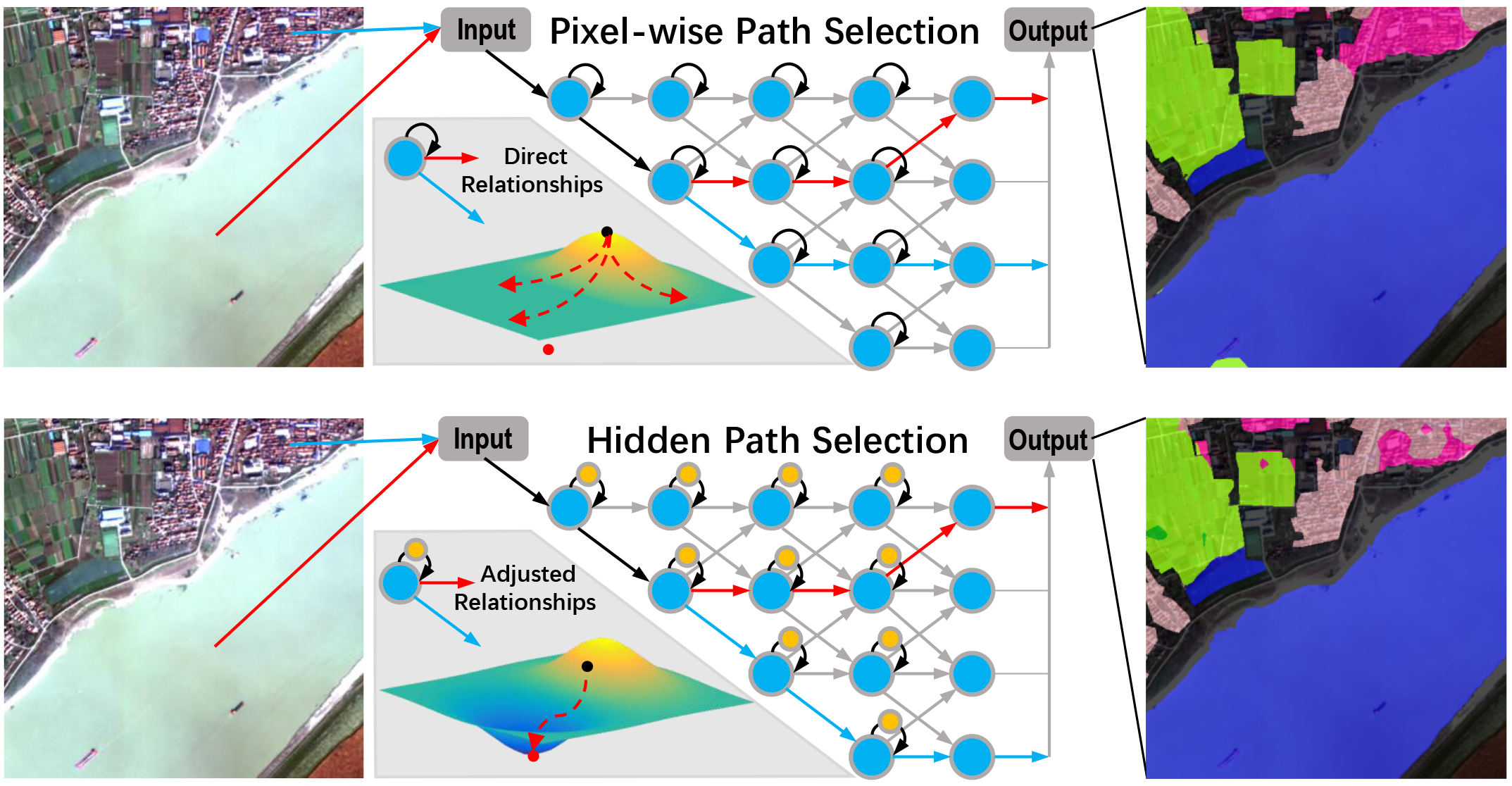}}
		\vspace{-3mm}
		\caption{\!\!\!In this paper, we provide  mathematical analyses of the parameter optimization in semantic segmentation of remote sensing images, which imply that direct relationships between feature maps and pixel-wise path selections in existing algorithms could make global optimums inaccessible. Consequently, we design the hidden path selection to adjust the aforementioned deleterious direct relationships through hidden variables and make the optimal path selections more possible to obtain. Straight lines with different colors represent selected paths for different pixels. Blue circles indicate the feature maps and orange circles indicate the hidden variables. }
		\label{Fig1}
		\vspace{-2mm}
	\end{figure} 	
	
	For implementing semantic segmentation, the methods to transfer the spectral signals into digital features first came into researchers' views, and classical classifiers in machine learning~\cite{1989An, 1947Regression} could be applied to predict pixel-wise land-cover categories \cite{2021Development, Min2005Decision, Erna2001Predicting}, where elaborate computation procedures could be exploited to obtain features corresponding to diverse land-cover distributions. Nevertheless, with pre-defined man-made calculations, hand-crafted features are hard to represent numberless different land-cover distributions in real-world environments, which has encouraged following researches to focus on improving the flexibility and representation capabilities of features through learnable deep structures~\cite{2020Land, 2020Deep, DBLP:conf/cvpr/YangLLX19}.
	
	However, although optimal parameters could be adapted to land-cover distributions existing in training data samples, the deep structures are often static after being learned in the training processes, which could lose the adaptability to complicated real-world environments during the inference procedures \cite{DBLP:conf/cvpr/LiSCLZWS20}. 
	In order to address this issue, many researchers have designed various elaborate structures to capture complex contextual information \cite{V3, V3+, PSPnet} not only for remote sensing images but also for natural ones. 
	
	Subsequently, neural architecture search (NAS) \cite{AutoDeepLab} is proposed to develop models with automatically acquired structures to adapt to diverse data distributions. 
	Furthermore, some researchers have proposed structures, called dynamic routing, to select forward paths adaptive to each data sample through designed activating factors \cite{DBLP:conf/cvpr/LiSCLZWS20} instead of adapting to data distributions of the whole dataset as in \cite{AutoDeepLab}.
	However, with all pixels sharing the same activating factors, such methods might ignore the diversity over different image areas, the effect of which is remarkable in remote sensing images due to the vast covered geographical areas with diverse land-cover distributions.
	Consequently, the Gated Path Selection Network (GPSNet) \cite{DBLP:journals/tip/GengZQHYZ21} has been proposed to select pixel-wise adaptive forward paths, simply shown at the top of Fig.\ref{Fig1}, through designed soft masks instead of same activating factors shared by all pixels, as in \cite{DBLP:conf/cvpr/LiSCLZWS20}. 
	However, considering all items in GPSNet as the variables, the direct functional relationships between soft masks (\textbf{executing pixel-wise path selections through multiplication}) and corresponding feature maps would make the parameter optimization constrained on a narrow high-dimension manifold instead of the full space, which might exclude the global optimums as illustrated in Fig.\ref{Fig1}.
	In other words, the direct relationships between feature maps and path selections could make it difficult to obtain the optimal pixel-wise forward path selections that are adaptive to different land-cover distributions. In summary, how to depict the land-cover distributions in complicated real-world environments for different pixels during the inference procedure has not been solved well by existing algorithms, which is of great interest to investigate. 
	
	\begin{figure*}[!t]
		\centering
		{\includegraphics[width=0.99\linewidth]{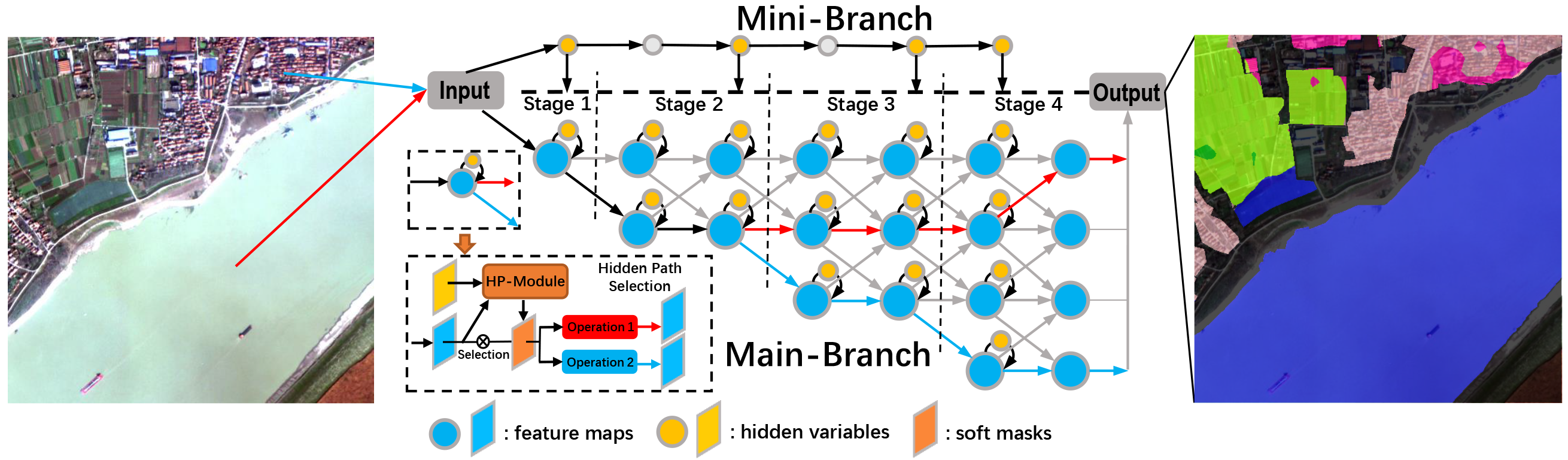}}
		\vspace{-3mm}
		\caption{
			{\em Hidden Path Selection Network} (HPS-Net) for semantic segmentation in remote sensing images. 
			HPS-Net utilizes an extra mini-branch to generate hidden variables for the main-branch, which can be utilized to obtain soft masks for selecting pixel-wise paths by multiplication in the main-branch through designed Hidden Path Module (HP-Module).
			Taking arbitrary deep structures as the main-branch, HP-Module integrates feature maps in the main-branch and corresponding hidden variables to calculate soft masks, which select paths from existing alternatives ({\eg} parallel structures or skip connections) in the structures of main-branch ({\eg} Resnet) for each pixel.
			In this paper, we take Resnet as the main-branch, while the mini-branch utilizes same structures with reduced channel numbers. We divide the mini-branch and main-branch into several stages according to the pixel resolution. 
			The last layer of each stage in mini-branch serves as hidden variables for the same stage in main-branch.
			In addition to the existing skip connections in Resnet, the last layer of each stage in main-branch would be considered as an extra alternative path for the following stages.}
		\label{Fig2}
		\vspace{-2mm}
	\end{figure*} 			
	
	In this paper, we portray the complicated pixel-wise land-cover distributions by imitating the hidden markov chain. As illustrated in Fig.\ref{Fig2}, we propose a method, called {\em Hidden Path Selection Network} (HPS-Net), with independent parallel structures ({\ie} a main-branch for semantic segmentation and a mini-branch for supporting) by exploiting a sequence of special modules called {\em Hidden Path Module} (HP-Module). Considering all items in the main-branch as variables, we notice that the high-dimension manifold where we execute the parameter optimization could be adjusted through utilizing hidden variables highly independent with main-branch. 
	Thus, we utilize the HP-Module to obtain a sequence of soft masks by leveraging hidden variables from the mini-branch to select pixel-wise adaptive forward paths in the main branch through multiplication, where the adjusted high-dimension manifold would have higher possibility to contain the global optimums.
	
	Further inspired by the mathematical analyses in terms of parameter optimization, we cut off the gradient flows from soft masks to the main deep structures during feature integration in the HP-Module to improve the training efficiency, while the gradient flows are retained intactly during the multiplication. Then, according to the Taylor expansion with respect to soft masks in HPS-Net, whose values are defined within a small range, the loss function can approximate a convex function to some extent if global optimums are contained in the aforementioned high-dimension manifold. 
	As a consequence, the usage of nearly freely valuing hidden variables would make soft masks adjusted freely and thus the convexity could help to obtain optimal soft masks (namely optimal pixel-wise path selections). As we shall see, with the proposed modules applied on existing deep structures, HPS-Net can better portray the pixel-wise land-cover distributions during the inference procedure.
	
	Another important aspect related to semantic segmentation in remote sensing images is the accessibility of well-annotated dataset. Although images in many existing datasets, {\eg \cite{2012The, Devis20172017, DBLP:journals/remotesensing/ZhangHZLLP17, 20182018}}, contain increasingly abundant land covers to compose different distributions, the included scenes can not perfectly reflect the complex real-world environments due to the limited data sample quantities, which would constrain the generalization abilities of trained deep models. Recently, a few large-scale datasets, {\eg~\cite{DBLP:conf/igarss/TongLXZ18, DeepGlobe18}}, have been built, but the involved land-cover categories are often insufficient yet and hardly meet the practical demands well. In this paper, we expand the land-cover classification dataset in \cite{2020Land} by refining the 5 involved land-cover categories in the whole GID-5 \cite{DBLP:conf/igarss/TongLXZ18} into 15 categories, which can better serve the practical demands.
	
	Our main contributions in this paper are threefold.
	\begin{itemize}
		\item We propose a new method, {{\ie} HPS-Net}, to imitate hidden markov chain, which depicts land-cover distributions and selects adaptive forward paths for each pixel through an extra mini-branch highly independent with arbitrary main deep structures.
		\item We provide mathematical analyses on the parameter optimization with taking all items in deep structures as variables to explain the inaccessibility of global optimums, guided by which we design detailed structures of the HPS-Net and propose the HP-Module to improve the possibility to obtain the global optimums.
		\item We expand the large-scale remote sensing semantic segmentation dataset, {{\ie} GID-5}, into 15 land-cover categories and propose the new GID-15 dataset, which provides a challenging large-scale benchmark platform with abundant land-cover categories and better satisfies the real-world scenarios.
		\vspace{3mm}
	\end{itemize}
	
	\section{Related Work}{\label{sec:related work}}
	\subsection{Semantic Segmentation}
	Early researches in remote sensing semantic segmentation often leveraged and developed the digital features, among which some works \cite{2021Development, Min2005Decision, Erna2001Predicting} rely on the physical properties of land-covers and some other works \cite{Ke2018Multi, Tuia2009Classification} capture the local spatial information. 
	Subsequently, many deep structures \cite{VGG, Resnet, GoogleNet, Alexnet}, whose fully connected layers are removed, have been borrowed in semantic segmentation models not only for remote sensing images but also for natural images \cite{FCN, segnet, Dconvnet}. 
	Numerous learnable parameters in deep structures can help obtain features with stronger representation capabilities. Especially, researches on depicting different object distributions, such as the existence of multi-scale objects, have been hot spots for a long time \cite{V3, V3+, PSPnet}. 
	Meanwhile, the community were inspired to research on neural architecture search (NAS) \cite{AutoDeepLab} for developing semantic segmentation models with automatically acquired structures instead of artificial structures. The architectural hyperparameters in these automatically acquired structures could make it more adaptive to different data distributions.
	
	Recently, some works have been proposed to focus on developing dynamic structures instead of static architectures (even automatically acquired). Specifically, activating factors obtained by designed soft conditional gates in \cite{DBLP:conf/cvpr/LiSCLZWS20} dynamically emphasize adaptive forward paths during the inference procedures. 
	However, land-cover distributions often vary significantly within one remote sensing image due to the wide range of covered areas. Thus, the diversity of land-cover distributions could require architectures to dynamically select adaptive forward paths for each pixel rather than for the whole image in \cite{DBLP:conf/cvpr/LiSCLZWS20}.
	
	In this paper, we propose the HPS-Net by imitating the hidden markov chain, where an extra mini-branch assists to better select pixel-wise adaptive forward paths in the main-branch for implementing semantic segmentation.

	\subsection{Contextual Information Modeling}
	In order to capture the contextual information, which implies the relationships between different stuffs, various basic units have been combined in different semantic segmentation structures. Specifically, pooling with different sizes \cite{PSPnet} and atrous convolutions with different dilation rates \cite{V3, V3+, dilation} are utilized to incorporate the contextual information of diverse stuffs with different scales. Meanwhile, skip connection structures have been designed to retain features of high resolutions for recovering detailed information \cite{Resnet, U-Net}.
	Furthermore, attention mechanisms have been designed in deep structures \cite{Asymmetric_Non-Local_Neural, PAN} to dynamically adjust contextual information for every pixel, where spatial attention and channel attention are aggregated. Especially, non-local operations were proposed to obtain dynamic attention maps through pixel-wise similarities \cite{DBLP:conf/cvpr/0004GGH18}.
	
	Recently, soft masks have been proposed \cite{DBLP:journals/tip/GengZQHYZ21} to adjust the contextual information, which can be seen as the combination of several normalized attention maps to dynamically select adaptive forward paths for every pixel. 
	However, according to the mathematical analyses in terms of parameter optimization we shall provide in this paper, the direct functional relationships between the soft masks and corresponding features in existing pixel-wise dynamic structures \cite{DBLP:journals/tip/GengZQHYZ21} would make it difficult to obtain the global optimums. 
	
	Consequently, we design an HP-Module to take the advantage of hidden variables derived from the extra mini-branch and adjust the aforementioned deleterious direct relationships, which makes HPS-Net more likely to obtain the optimal pixel-wise path selections.

	\subsection{Benchmark Dataset for Semantic Segmentation}
	To support developing semantic segmentation algorithms in remote sensing images, benchmark datasets are important to serve as the basis for training and evaluation.
	Widely used datasets \cite{2012The}, {{\eg} {\em ISPRS Test Project on Urban Classification and 3D Building Reconstruction}}, focuses on urban areas in cities such as Vaihingen. Some contests on other topics, such as {\em IEEE GRSS Data Fusion Contest} in 2017 and 2018, also provide semantic segmentation datasets \cite{Devis20172017, 20182018} covering several cities, while \cite{DBLP:journals/remotesensing/ZhangHZLLP17} provides data samples from different platforms. 
	Besides, some datasets concentrating on specific categories, such as buildings and roads, are also provided \cite{2017Learning}. 
	Further to expand the quantity of data samples, several large scale datasets have been created \cite{DBLP:conf/igarss/TongLXZ18, DeepGlobe18}, which cover larger geographical areas to ensure the diversity of data samples. However, existing datasets hardly give considerations to both the quantity of images and sufficiency of land-cover categories. For instance, although GID-5 \cite{DBLP:conf/igarss/TongLXZ18} contains a large quantity of data samples and covers wide ranges of geographical areas, the limited 5 land-cover categories are often insufficient for real-world applications.
	
	In this paper, we expand the GID-5 by refining 5 general categories into 15 elaborated categories and create the GID-15 dataset, which can serve as a challenging benchmark and better meet the demands in real-world applications.

	\section{Methods}\label{Method}
	\subsection{Mathematical Formulation}\label{Math_Modelization}
	Interpreting the image ${\small{\mathbf{I}}}$ of $\mathrm{c} \in \mathbb{N}_+$ channels as a mapping function belongs to a special function family $\mathscr{F}$, we have ${\small{\mathbf{I}}}:\Omega \rightarrow \mathbb{R}^{\mathrm{c}}$ defined on the image grid $\Omega: \{0,1,\ldots,H-1\} \times \{0,1,\ldots,W-1\}$, which would return values of $\mathrm{c}$ channels by given a coordinate of position $\mathbf p$ in $\Omega$. Consequently, the semantic segmentation model can be interpreted as a mapping function that
	\begin{equation}\label{eq1}
	\forall \ \ {\small{\mathbf{I}}} \in \mathscr{F}, \ \mathbf p \in \Omega, \ f(\mathbf{I}, \mathbf p) = \arg\max_{\mathrm{l_k} \in \mathbf{L}}{\mathcal{M}_{\mathbf{I}}({\mathbf p})}.
	\end{equation}
	$\mathbf{L}=\{\mathrm{l_k}\}_{\mathrm{k} \in \mathbb{N}_+ \le \mathrm{N}}$ is a label set containing $\mathrm{N}$ categories and probability map $\mathcal{M}_{\mathbf{I}}$ measures the probability of each category corresponding to each position $\mathbf p \in \Omega$.
	
	The structures of conventional deep models often combine various elaborate units, such as convolutions and nonlinear activation functions, to capture complex contextual information adaptive to various distributions. Therefore, considering each layer in existing deep structures, such as \cite{Resnet}, as a basic function consisting of different units, the whole structures of $\mathrm{D} \in \mathbb{N}_+$ layers can be interpreted as a sequence of nested composite functions that
	\begin{equation}\label{eq2}
	\begin{split}
	\forall \ \ \mathbf{F}_{1}=& \ \ \mathbf{I}, \ \ \mathrm{d} \in \mathbb{N}_+ < \mathrm{D}, \ \ \mathbf{F}_{\mathrm{d+1}} = f_\mathrm{d} (\{\mathbf{F}_{\mathrm{i}}\}_{\mathrm{i} \in \mathbb{N}_+ \le \mathrm{d}}),\\
	&\mathcal{M}_{\mathbf{I}}=\mathbf{F}_{\mathrm{D}}, \ \ \ f(\mathbf{I}, \mathbf p) = \arg\max_{\mathrm{l_k} \in \mathbf{L}}{\mathcal{M}_{\mathbf{I}}({\mathbf p})}.
	\end{split}
	\end{equation}
	$f_\mathrm{d}$ represents the basic function in the ${\mathrm{d}}$-th layer and $\mathbf{F}_{\mathrm{d}}$ represents the corresponding feature maps. Considering in the skip connection structures, $f_\mathrm{d}$ receives feature maps $\{\mathbf{F}_{\mathrm{i}}\}_{\mathrm{i} \in \mathbb{N}_+ \le \mathrm{d}}$ in all the previous layers (not all necessarily participate in calculations) and returns the feature maps $\mathbf{F}_{\mathrm{d+1}}$ for $f_\mathrm{d+1}$ in the next layer. Consequently, the first layer in deep structures receives image $\mathbf{I}$ as the input and feature map $\mathbf{F}_{\mathrm{D}}$ in the last layer serves as the probability map $\mathcal{M}_{\mathbf{I}}$ to measure the pixel-wise probability of each category. It is worth noticing that models derived from NAS can also be interpreted in this form.
	
	In order to better adapt to complicated real-world scenarios, dynamic routing \cite{DBLP:conf/cvpr/LiSCLZWS20} is proposed to select image-dependent forward paths in deep structures. Specifically, we can interpret dynamic routing as 
	\begin{equation}\label{eq3}
	\begin{split}
	\forall \ \mathbf{F}_{1}=& \ \mathbf{I}, \ \mathrm{d} \in \mathbb{N}_+ < \mathrm{D}, \ \mathbf{F}_{\mathrm{d+1}} = f_\mathrm{d} (\{\mathbf{F}_{\mathrm{i}}\}_{\mathrm{i} \in \mathbb{N}_+ \le \mathrm{d}}, \alpha^{\mathrm{d}}),\\
	&\mathcal{M}_{\mathbf{I}}=\mathbf{F}_{\mathrm{D}}, \ \ \ f(\mathbf{I}, \mathbf p) = \arg\max_{\mathrm{l_k} \in \mathbf{L}}{\mathcal{M}_{\mathbf{I}}({\mathbf p})},
	\end{split}
	\end{equation}	
	where $\alpha^{\mathrm{d}} \in \mathbb{R}^{s}$ represents the activation factor for total $s \in \mathbb{N}_+$ forward paths in the $\mathrm{d}$-th layer. In the implementation, dynamic routing utilizes learnable structures to compose the mapping function $\mathbf{G}_{\mathrm{d}} : \mathbf{I} \rightarrow \mathbb{R}^{3}$ and obtain $\alpha^{\mathrm{d}}$ for selecting forward paths corresponding to up-scale, equivalent-scale and down-scale. The activation factor would be multiplied with corresponding feature maps to select paths adaptive to different scale distributions among the whole input image.
	
	\begin{figure}[!t]
		\centering
		{\includegraphics[width=1\linewidth]{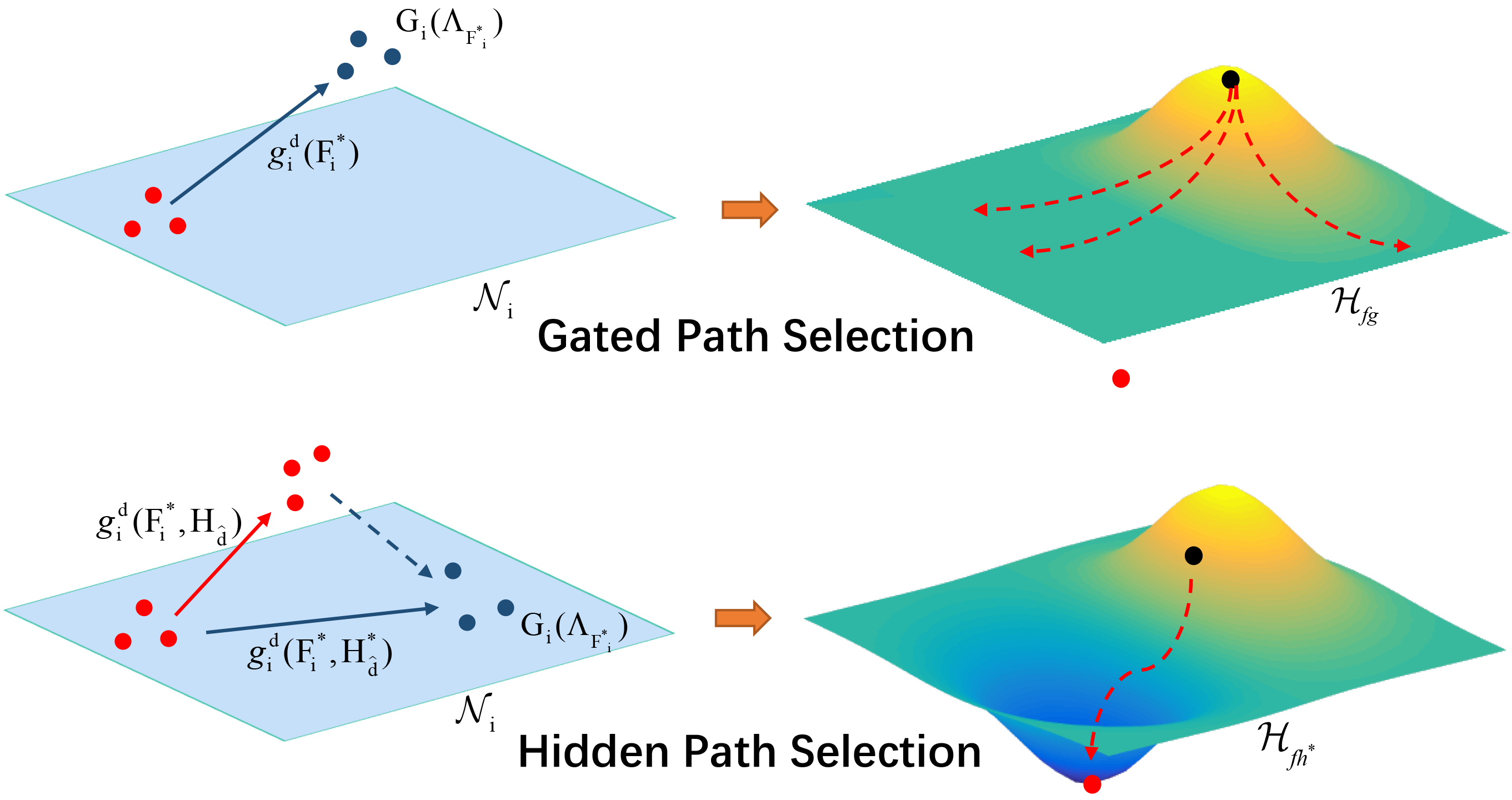}}
		\vspace{-3mm}
		\caption{In gated path selection, $g^{\mathrm{d}}_{\mathrm{i}}$ is hard to meet the required property that numbers of specific points in $\mathcal{N}_\mathrm{i}$ should be mapped within $\mathcal{N}_\mathrm{i}$ through $g^{\mathrm{d}}_{\mathrm{i}}$ ({\ie} $\mathrm{G}_{\mathrm{i}}$), which could exclude the global optimums (red point) outside the manifold $\mathcal{H}_{fg}$ for executing the parameter optimization. In hidden path selection, $g^{\mathrm{d}}_{\mathrm{i}}$ could be assisted by changing different $\mathbf{H}_{\hat{\mathrm{d}}}$ and thus the manifold $\mathcal{H}_{fh}$ could be adjusted to approach the global optimums. Especially, $\mathbf{H}^{*}_{\hat{\mathrm{d}}}$ and $\mathcal{H}^{*}_{fh}$ represent the optimal $\mathbf{H}_{\hat{\mathrm{d}}}$ and $\mathcal{H}_{fh}$ we obtain respectively. } 
		\label{Fig3}
		\vspace{-2mm}
	\end{figure} 
	
	\subsection{Gated Path Selection}\label{Gated_Path_Selection}	
	Although dynamic routing mentioned in Sec.\ref{Math_Modelization} can adapt to scale distributions during the inference procedure, it rarely considers the diversity over different image areas, which is remarkable in remote sensing images due to the wide range of covered geographical areas. Recently proposed GPSNet \cite{DBLP:journals/tip/GengZQHYZ21} extends the idea of dynamic routing, where the utilized soft masks can be seen as the activation factor maps to select adaptive forward paths for each pixel. Specifically, we have 
	\begin{equation}\label{eq4}
	\begin{split}
	\forall \ \mathbf{F}_{1}=& \ \mathbf{I}, \mathrm{d} \in \mathbb{N}_+ < \mathrm{D} \ , \mathbf{F}_{\mathrm{d+1}} = f_\mathrm{d} (\{\mathbf{F}_{\mathrm{i}} \odot \mathbf{W}^{\mathrm{d}}_{\mathrm{i}}\}_{\mathrm{i} \in \mathbb{N}_+ \le \mathrm{d}}),\\
	&\mathcal{M}_{\mathbf{I}}=\mathbf{F}_{\mathrm{D}}, \ \ \ f(\mathbf{I}, \mathbf p) = \arg\max_{\mathrm{l_k} \in \mathbf{L}}{\mathcal{M}_{\mathbf{I}}({\mathbf p})}.
	\end{split}
	\end{equation}	
	$\mathbf{W}^{\mathrm{d}}_{\mathrm{i}}$ is the soft mask of the same spatial size with feature map $\mathbf{F}_{\mathrm{i}}$, where the element in each position serves as a probability for pixel-wise forward path selections. The symbol $\odot$ represents the element-wise product. It is worth noticing that $\mathbf{W}^{\mathrm{d}}$ in the traditional layers can be seen as tensors full with $\mathbf{1}$. In the implementation, GPSNet directly utilizes $\mathbf{F}_{\mathrm{i}}$ to generate $\mathbf{W}^{\mathrm{d}}_{\mathrm{i}}$, namely $\mathbf{W}^{\mathrm{d}}_{\mathrm{i}} = g^{\mathrm{d}}_{\mathrm{i}}(\mathbf{F}_{\mathrm{i}})$ highly dependent with $\mathbf{F}_{\mathrm{i}}$ for each $\mathrm{i}$ with $g^{\mathrm{d}}_{\mathrm{i}}$ being a differentiable function, which would \textbf{make it difficult to obtain the global optimums of the problem as we analyze in the following texts}.
	
	Noticing that with features in deep structures being considered as variables determined by parameter $\bm{\theta}_f$ in $\{f_\mathrm{d}\}_{\mathrm{d} \in \mathbb{N}_+ < \mathrm{D}}$, the optimization of $\bm{\theta}_f$ can be seen as searching the global optimums within a high-dimension manifold $\mathcal{H}_f$ in the full space, whose shape is determined by the structures we design for $\{f_\mathrm{d}\}_{\mathrm{d} \in \mathbb{N}_+ < \mathrm{D}}$.  
	Moreover, optimizing parameters $\bm{\theta}_f$ in all $\{f_\mathrm{d}\}_{\mathrm{d} \in \mathbb{N}_+ < \mathrm{D}}$ during the training process would guide us to find $\mathbf{F^*}_{\mathrm{i}}$ and $\mathbf{W}^{\mathrm{*d}}_{\mathrm{i}}$ for each $\mathbf{I}$ determined by the optimal parameters. Starting from this view, we calculate $\frac{\partial L}{\partial \mathbf{F^*}_{\mathrm{i}}}$ and $\frac{\partial L}{\partial \mathbf{W*}^{\mathrm{d}}_{\mathrm{i}}}$ with $L$ as the loss function such that
	\begin{equation}\label{eq5}
	\begin{split}
	&\frac{\partial L}{\partial \mathbf{F^*}_{\mathrm{i}}}= \sum_{\mathrm{d}<\mathrm{D}} (\frac{\partial L}{\partial f_\mathrm{d}} \bigcdot \frac{\partial f_\mathrm{d}}{\partial \mathbf{F^*}_{\mathrm{i}} \odot  \mathbf{W}^{\mathrm{*d}}_{\mathrm{i}}}) \bigcdot  \mathbf{W}^{\mathrm{*d}}_{\mathrm{i}}, \ \ \\
	&\forall \mathrm{d}<\mathrm{D}, \ \frac{\partial L}{\partial \mathbf{W}^{\mathrm{*d}}_{\mathrm{i}}}= \frac{\partial L}{\partial f_\mathrm{d}} \bigcdot \frac{\partial f_\mathrm{d}}{\partial \mathbf{F^*}_{\mathrm{i}} \odot  \mathbf{W}^{\mathrm{*d}}_{\mathrm{i}}} \bigcdot \mathbf{F^*}_{\mathrm{i}},
	\end{split} 	
	\end{equation}		
	where $\bigcdot$ represents the matrix product.	Importantly, $\frac{\partial L}{\partial \mathbf{F^*}_{\mathrm{i}}}$ and $\frac{\partial L}{\partial \mathbf{W}^{\mathrm{*d}}_{\mathrm{i}}}$ \textbf{should be null tensor $\mathbf{0}$ to retain the probability that $\mathbf{F^*}_{\mathrm{i}}$ and $\mathbf{W}^{\mathrm{*d}}_{\mathrm{i}}$ equal global optimums, which are denoted as $\mathbf{F^{\circ}}_{\mathrm{i}}$ and $\mathbf{W}^{\mathrm{{\circ}d}}_{\mathrm{i}}$, and $\mathcal{H}_f$ contains the global optimums.} 
	
	Thus, further with the direct functional relationships $\mathbf{W}^{\mathrm{d}}_{\mathrm{i}} = g^{\mathrm{d}}_{\mathrm{i}}(\mathbf{F}_{\mathrm{i}})$, we require two formulas tenable that
	\begin{equation}\label{eq6}
	\begin{split}
	&\sum_{\mathrm{d}<\mathrm{D}} (\frac{\partial L}{\partial f_\mathrm{d}} \bigcdot \frac{\partial f_\mathrm{d}}{\partial \mathbf{F^*}_{\mathrm{i}} \odot  \mathbf{W}^{\mathrm{*d}}_{\mathrm{i}}}) \bigcdot  g^{\mathrm{d}}_{\mathrm{i}}(\mathbf{F}_{\mathrm{i}}) =  \Lambda_{\mathrm{i}} \bigcdot \mathrm{G}_{\mathrm{i}}(\Lambda_{\mathbf{F^*}_{\mathrm{i}}}) = \mathbf{0}, \ \ \\
	&\sum_{\mathrm{d}<\mathrm{D}} (\frac{\partial L}{\partial f_\mathrm{d}} \bigcdot \frac{\partial f_\mathrm{d}}{\partial \mathbf{F^*}_{\mathrm{i}} \odot  \mathbf{W}^{\mathrm{*d}}_{\mathrm{i}}}) \bigcdot \mathbf{F^*}_{\mathrm{i}} = \Lambda_{\mathrm{i}} \bigcdot \Lambda_{\mathbf{F^*}_{\mathrm{i}}} = \mathbf{0}.
	\end{split} 	
	\end{equation}	
	$\Lambda_{\mathrm{i}}$ and $\Lambda_{\mathbf{F^*}_{\mathrm{i}}}$ are block matrices that list all product factors in the summation on the left side of Eq.\eqref{eq6}, while function $\mathrm{G}_{\mathrm{i}}$ is totally determined by $g^{\mathrm{d}}_{\mathrm{i}}$. Consequently, Eq.\eqref{eq6} illustrates that the column vectors of $\Lambda_{\mathbf{F^*}_{\mathrm{i}}}$ and $\mathrm{G}_{\mathrm{i}}(\Lambda_{\mathbf{F^*}_{\mathrm{i}}})$ should be all within the null space $\mathcal{N}_\mathrm{i}$ of $\Lambda_{\mathrm{i}}$, which requires \textbf{numbers of specific points in $\mathcal{N}_\mathrm{i}$ to be mapped within $\mathcal{N}_\mathrm{i}$ through $\mathrm{G}_{\mathrm{i}}$}.
	In other words, the optimization of $\bm{\theta}_f$ can be seen within a manifold $\mathcal{H}_{fg} \subset \mathcal{H}_{f}$, which is narrow down, due to the extra functional relationships $\mathbf{W}^{\mathrm{d}}_{\mathrm{i}} = g^{\mathrm{d}}_{\mathrm{i}}(\mathbf{F}_{\mathrm{i}})$. Then, the extra constraint illustrated above caused by Eq.\eqref{eq6} may not be satisfied by optimal $\bm{\theta}_f$, which means the narrow $\mathcal{H}_{fg}$ could exclude the global optimums.

	
	Although $\mathcal{N}_\mathrm{i}$ is a linear space, nonlinear units such as softmax or sigmoid are always utilized to fix the ranges of $\mathrm{G}_{\mathrm{i}}(\Lambda_{\mathbf{F^*}_{\mathrm{i}}})$.
	Thus, as illustrated at the top of Fig.\ref{Fig3}, the designed $g^{\mathrm{d}}_{\mathrm{i}}$, which determines $\mathrm{G}_{\mathrm{i}}$, is hard to meet the required property ({\ie} meeting Eq.\eqref{eq6}) without the prior knowledge of $\mathcal{N}_\mathrm{i}$. Consequently, the narrowed manifold $\mathcal{H}_{fg}$ could exclude mentioned global optimums no matter how elaborate the $g^{\mathrm{d}}_{\mathrm{i}}$ is. 
	The same problem would also exist when $\mathbf{W}^{\mathrm{*d}}_{\mathrm{i}}$ has functional relationships with previous layers involved in the calculations of $\mathbf{F^*}_{\mathrm{i}}$, because it also requires $g^{\mathrm{d}}_{\mathrm{i}}$ to meet the additional property and all restrictions could only be released by optimizing the parameters $\bm{\theta}_f$.

	\subsection{Hidden Path Selection}\label{Hidden_Path_Selection}
	In order to release the restriction of $g^{\mathrm{d}}_{\mathrm{i}}$ caused by direct dependences between $\mathbf{F}_{\mathrm{i}}$ and $\mathbf{W}^{\mathrm{d}}_{\mathrm{i}}$ analyzed in Sec.\ref{Gated_Path_Selection}, we naturally introduce extra variables $\mathbf{H}_{\mathrm{d}}$ highly independent with $\mathbf{F}_{\mathrm{i}}$ to participate in the calculations in $g^{\mathrm{d}}_{\mathrm{i}}$, namely $\mathbf{W}^{\mathrm{d}}_{\mathrm{i}} = g^{\mathrm{d}}_{\mathrm{i}}(\mathbf{F}_{\mathrm{i}}, \mathbf{H}_{\mathrm{d}})$. Serving as the probability mask to indicate the adaptive forward paths for each pixel in $\mathbf{F}_{\mathrm{i}}$, $\mathbf{W}^{\mathrm{d}}_{\mathrm{i}}$ is naturally decided by the properties of different areas in the input image $\mathbf{I}$. Thus, we design the $\mathbf{H}_{\mathrm{d}}$ as the mapping function on $\mathscr{F}$ defined in Sec.\ref{Math_Modelization} that takes $\mathbf{I}$ as the input. Specifically, we have 
	\begin{equation}\label{eq7}
	\mathbf{W}^{\mathrm{d}}_{\mathrm{i}} = g^{\mathrm{d}}_{\mathrm{i}}(\mathbf{F}_{\mathrm{i}}, \mathbf{H}_{\mathrm{d}}), \ \ \mathbf{H}_{\mathrm{d}}: \mathscr{F} \rightarrow \mathbb{R}^{\mathrm{s}_\mathrm{d}},
	\end{equation}	
	where $\mathrm{s}_\mathrm{d}$ is a positive integer vector which indicates the shape of $\mathbf{H}_{\mathrm{d}}$. With the help of $\mathbf{H}_{\mathrm{d}}$, the structures of $g^{\mathrm{d}}_{\mathrm{i}}$ could be of no restrictions to some extent, where $\mathbf{H}_{\mathrm{d}}$ can adjust $\mathrm{G}_{\mathrm{i}}(\Lambda_{\mathbf{F^*}_{\mathrm{i}}})$ to approach $\mathcal{N}_\mathrm{i}$ for each $\Lambda_{\mathbf{F^*}_{\mathrm{i}}}$ in $\mathcal{N}_\mathrm{i}$. Equivalently, the shape of manifold $\mathcal{H}_{fg}$ would be adjusted by $\mathbf{H}_{\mathrm{d}}$ to approach the global optimums. Consequently, the global optimal with features in deep structures considered as variables are more likely to be searched during the optimization of $\bm{\theta}_f$.
	
	Actually, $\mathbf{H}_{\mathrm{d}}$ can be seen as the hidden variables that are unobservable solely given $\mathbf{F}_{\mathrm{i}}$, which inspires us to imitate the hidden markov chain. Specifically, we design a light-weight mini-branch, which generates $\mathbf{H}_{\mathrm{d}}$ to guide forward paths for each pixel of $\mathbf{F}_{\mathrm{i}}$ in the main-branch through Eq.\eqref{eq4} and Eq.\eqref{eq7}, namely \textbf{hidden path selection}.
	
	
	\begin{algorithm}\label{algorithm}
		\caption{Hidden Path Selection Network}
		\LinesNumbered
		
		\KwIn{image $\mathbf{I}$, $\mathscr{L}_{\mathrm{d}}=\{\mathrm{d}_\mathrm{k}\}_{k \in \mathbb{N}_+}$ for the main-branch, $\mathscr{H}_{\mathrm{d}}=\{\mathrm{d}_\mathrm{m}\}_{m \in \mathbb{N}_+}$ for the mini-branch}
		
		\KwOut{Semantic Segmentation Results}
		
		$\mathbf{H}_{1}=\mathbf{I}, \mathbf{F}_{1}=\mathbf{H}_{1}$\;
		
		\For{$\bar{\mathrm{d}} = 1; \bar{\mathrm{d}} < \mathrm{D}$}
		{
			$\mathbf{H}_{\bar{\mathrm{d}}+1} = \  h_{\bar{\mathrm{d}}} (\{\mathbf{H}_{\mathrm{i}}\}_{\mathrm{i} \in \mathbb{N}_+ \le \bar{\mathrm{d}}})$\;
		}
		
		
		\For{$\mathrm{d} = 1; \mathrm{d} < \mathrm{D}$}
		{
			$\hat{\mathrm{d}} = \mathrm{min}\{\mathrm{d}_\mathrm{m} \in \mathscr{H}_{\mathrm{d}}, \mathrm{d} \leq \mathrm{d}_\mathrm{m}\}$\;
			
			\If{ $\mathrm{d} \in \mathscr{L}_{\mathrm{d}}$}
			{
				$\mathbf{F}_{\mathrm{d+1}} = f_\mathrm{d} (\{\mathbf{F}_{\mathrm{i}} \odot g^{\mathrm{d}}_{\mathrm{i}}(\mathbf{F}_{\mathrm{i}}, \mathbf{H}_{\hat{\mathrm{d}}})\}_{\mathrm{i} \in \mathbb{N}_+ \le \mathrm{d}})$\;
			}
			\Else{
				$\mathbf{F}_{\mathrm{d+1}} = f_\mathrm{d} (\{\mathbf{F}_{\mathrm{i}}\}_{\mathrm{i} \in \mathbb{N}_+ \le \mathrm{d}})$\;
			}
		}
		
		$\mathcal{M}_{\mathbf{I}}=\mathbf{F}_{\mathrm{D}}$\;
		
		\textbf{final}\;
		
		\textbf{return} $\arg\max_{\mathrm{l_k} \in \mathbf{L}}{\mathcal{M}_{\mathbf{I}}({\mathbf p})}, \ \forall \ \mathbf p \in \Omega$;
	\end{algorithm}

	\subsection{Hidden Path Selection Network}\label{Hidden_Selection_Network}
	As illustrated in Fig.\ref{Fig2}, in the proposed HPS-Net, we can implement hidden path selection on existing deep structures ({\eg} Resnet-101), which serve as the main branch, based on the analyses in Sec.\ref{Gated_Path_Selection} and Sec.\ref{Hidden_Path_Selection}. 
	An extra mini-branch consisting of the same structure as main-branch with all channel numbers reduced to 32 is utilized to generate hidden variables $\mathbf{H}_{\mathrm{d}}$ in Eq.\eqref{eq7}.
	Specifically, we utilize a layer-index set $\mathscr{L}_{\mathrm{d}}=\{\mathrm{d}_\mathrm{k}\}_{k \in \mathbb{N}_+}$ for indicating which layer in the main-branch to be applied hidden path selection. Denoting the basic functions in each layer of mini-branch as $\{h_{\bar{\mathrm{d}}}\}_{\bar{\mathrm{d}} \in \mathbb{N}_+ < \mathrm{D}}$, we also utilize a layer-index set $\mathscr{H}_{\mathrm{d}}=\{\mathrm{d}_\mathrm{m}\}_{m \in \mathbb{N}_+}$ to indicate which feature maps in the mini-branch to serve as hidden variables for hidden path selection. Concretely, we modelize the HPS-Net as
	\begin{equation}\label{eq8}
	\begin{split}
	&\forall \ \mathbf{H}_{1}=\mathbf{I}, \ \ \bar{\mathrm{d}} \in \mathbb{N}_+ < \mathrm{D}, \ \ \mathbf{H}_{\bar{\mathrm{d}}+1} = \  h_{\bar{\mathrm{d}}} (\{\mathbf{H}_{\mathrm{i}}\}_{\mathrm{i} \in \mathbb{N}_+ \le \bar{\mathrm{d}}}),\\
	&\forall \ \mathbf{F}_{1}=\mathbf{H}_{1}, \ \mathrm{d} \in \mathbb{N}_+ < \mathrm{D}, \ \hat{\mathrm{d}} = \mathrm{min}\{\mathrm{d}_\mathrm{m} \in \mathscr{H}_{\mathrm{d}}, \mathrm{d} \leq \mathrm{d}_\mathrm{m}\},\\
	&\mathbf{F}_{\mathrm{d+1}} = \left\{\begin{array}{ll}
	f_\mathrm{d} (\{\mathbf{F}_{\mathrm{i}} \odot g^{\mathrm{d}}_{\mathrm{i}}(\mathbf{F}_{\mathrm{i}}, \mathbf{H}_{\hat{\mathrm{d}}})\}_{\mathrm{i} \in \mathbb{N}_+ \le \mathrm{d}}), & \text{if}\,~\mathrm{d} \in \mathscr{L}_{\mathrm{d}},\\
	f_\mathrm{d} (\{\mathbf{F}_{\mathrm{i}}\}_{\mathrm{i} \in \mathbb{N}_+ \le \mathrm{d}}), &
	\text{otherwise},
	\end{array}\right.,\\
	&\mathcal{M}_{\mathbf{I}}=\mathbf{F}_{\mathrm{D}}, \ \ \ f(\mathbf{I}, \mathbf p) = \arg\max_{\mathrm{l_k} \in \mathbf{L}}{\mathcal{M}_{\mathbf{I}}({\mathbf p})},
	\end{split}
	\end{equation}	
	where the output $\mathbf{F}_{\mathrm{D}}$ of main-branch finally implements semantic segmentation. The detail process can be found in Algorithm.\ref{algorithm}.
	
\begin{figure}[!t]
		\centering
		{\includegraphics[width=0.8\linewidth]{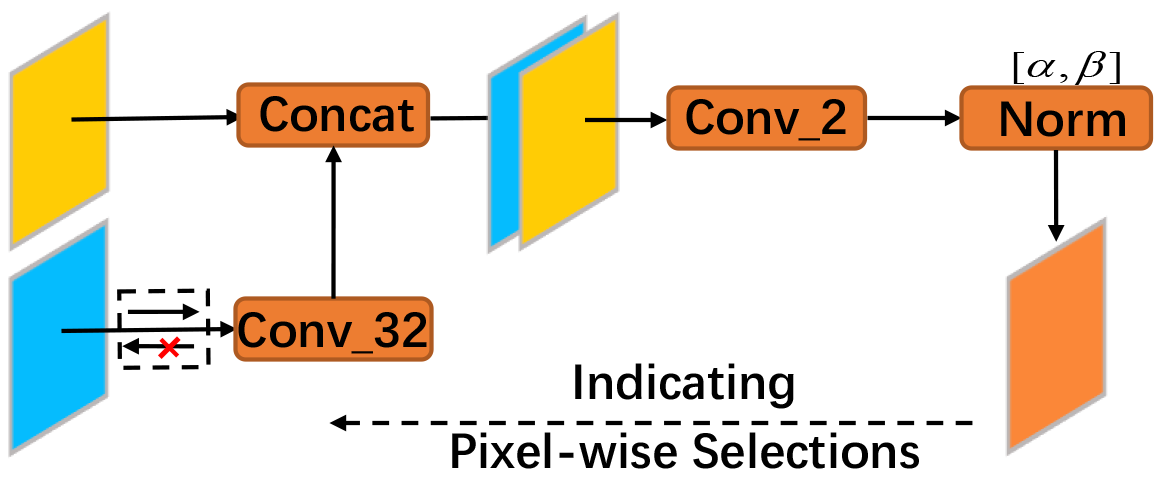}}
		\vspace{-3mm}
		\caption{The proposed HP-Module first utilizes convolutions with channel numbers of 32 and concatenations to integrate the feature maps from main-branch and hidden variables derived from mini-branch. Then, the convolution with channel numbers of 2 and normalization functions generate the soft masks with values constrained between $[\alpha, \beta]$, which is utilized to indicate the pixel-wise path selections for the feature maps in the main-branch.} 
		\label{Fig4}
		\vspace{-2mm}
	\end{figure}

	In the implementation, as illustrated in Fig.\ref{Fig4}, we propose {\em{Hidden Path Module}} (HP-Module) to generate $\mathbf{W}^{\mathrm{d}}_{\mathrm{i}}=g^{\mathrm{d}}_{\mathrm{i}}(\mathbf{F}_{\mathrm{i}}, \mathbf{H}_{\hat{\mathrm{d}}})$, which consists of feature concatenations, convolutions and normalization functions. We have
	\begin{equation}\label{eq9}
	\mathbf{W}^{\mathrm{d}}_{\mathrm{i}}=[f^{\alpha, \beta}_{norm}(f_{conv\_2}(f_{cat}(f_{conv\_32}(\mathbf{F}_{\mathrm{i}}), \mathbf{H}_{\hat{\mathrm{d}}})))]_{\mathrm{i}}.
	\end{equation}		
	$f_{conv\_2}$ and $f_{conv\_32}$ are convolutions with 2 and 32 output channels respectively, where kernel sizes are all set as $3 \times 3$. $f_{cat}$ is the concatenation operation. $f^{\alpha, \beta}_{norm}$ is the normalization function consisting of softmax and clip function, which normalizes the values within $[\alpha, \beta]$. Importantly, $[ \ ]_{\mathrm{i}}$ indicates the slice alongside the channel dimension according to index $\mathrm{i}$, which is corresponding to the i-th alternative path. 
	
	Recalling analyses on the global optimums in Sec.\ref{Gated_Path_Selection}, we further calculate the gradient flows of $\mathbf{F}_{\mathrm{i}}$ with $\mathbf{W}^{\mathrm{d}}_{\mathrm{i}}=g^{\mathrm{d}}_{\mathrm{i}}(\mathbf{F}_{\mathrm{i}}, \mathbf{H}_{\hat{\mathrm{d}}})$ that
	\begin{equation}\label{eq10}
	\frac{\partial L}{\partial \mathbf{F}_{\mathrm{i}}}=\sum_{\mathrm{d}<\mathrm{D}} \frac{\partial L}{\partial f_\mathrm{d}} \bigcdot \frac{\partial f_\mathrm{d}}{\partial \mathbf{F}_{\mathrm{i}} \odot  \mathbf{W}^{\mathrm{d}}_{\mathrm{i}}} \bigcdot (\mathbf{W}^{\mathrm{d}}_{\mathrm{i}}+\mathbf{F}_{\mathrm{i}} \bigcdot \frac{\partial \mathbf{W}^{\mathrm{d}}_{\mathrm{i}}}{\partial \mathbf{F}_{\mathrm{i}}}).
	\end{equation}	
	
	It is worth noticing that Eq.\eqref{eq5} and Eq.\eqref{eq10} are not contradicted. Eq.\eqref{eq5} calculates the partial derivative of $\mathbf{F}_{\mathrm{i}}$ with features in deep structures ({\eg} $\mathbf{W}^{\mathrm{d}}_{\mathrm{i}}$ and $\mathbf{F}_{\mathrm{i}}$) considered as variables, which is utilized to analyze the global optimums in the actual full space and not involved in the calculations among parameter optimization. Eq.\eqref{eq10} calculates the gradient flows in our proposed HPS-Net to optimize the parameter $\bm{\theta}_f$, which can be seen as searching parameters within a manifold $\mathcal{H}_{fh}$ adjusted by $\mathbf{H}_{\hat{\mathrm{d}}}$ based on $\mathcal{H}_{fg}$. Moreover, the efficiency of training process would be reduced with $\mathbf{W}^{\mathrm{d}}_{\mathrm{i}}$ being contrary sign to $\mathbf{F}_{\mathrm{i}} \bigcdot \frac{\partial \mathbf{W}^{\mathrm{d}}_{\mathrm{i}}}{\partial \mathbf{F}_{\mathrm{i}}}$, which is hard to avoid with constantly changing parameters $\bm{\theta}_f$ during the training process. Consequently, we cut-off the gradient flows from soft masks $\mathbf{W}^{\mathrm{d}}_{\mathrm{i}}$ to feature maps $\mathbf{F}_{\mathrm{i}}$ in the main-branch by multiplying null tensor $\mathbf{0}$ with the $\mathbf{F}_{\mathrm{i}} \bigcdot \frac{\partial \mathbf{W}^{\mathrm{d}}_{\mathrm{i}}}{\partial \mathbf{F}_{\mathrm{i}}}$ on the right side of Eq.\eqref{eq10}. As we shall see, by controlling gradient flows, the efficiency of training process would be increased.

	\subsection{Model Analyses}\label{Model_Analyses}
	To hark back to the manifold for parameter optimization, as illustrated at the bottom of Fig.\ref{Fig3}, the shape of $\mathcal{H}_{fh}$ could be adjusted by $\mathbf{H}_{\hat{\mathrm{d}}}$ based on $\mathcal{H}_{fg}$, where we realize that the high independence between $\mathbf{F}_{\mathrm{i}}$ and $\mathbf{H}_{\hat{\mathrm{d}}}$ makes it possible to let $\mathbf{H}_{\hat{\mathrm{d}}}$ value freely and adjust $\mathbf{W}^{\mathrm{d}}_{\mathrm{i}}=g^{\mathrm{d}}_{\mathrm{i}}(\mathbf{F}_{\mathrm{i}}, \mathbf{H}_{\mathrm{d}})$ ({\ie} adjust $\mathrm{G}_{\mathrm{i}}(\Lambda_{\mathbf{F^*}_{\mathrm{i}}})$) to meet the demanded property. Consequently, the larger range domain of $\mathbf{H}_{\hat{\mathrm{d}}}$ in the mini-branch would make $\mathcal{H}_{fh}$ adjusted more freely with the designed formulation of $g^{\mathrm{d}}_{\mathrm{i}}$, where $\mathcal{H}_{fh}$ is more likely to contain the global optimums with all feature maps considered as variables. 
	
	Moreover, the floating range of values in $\mathbf{W}^{\mathrm{d}}_{\mathrm{i}}$ is defined within $[\alpha, \beta]$, which is ensured by applying $f^{\alpha, \beta}_{norm}$. 
	Recalling that we denote the global optimums $\mathbf{W}^{\mathrm{d}}_{\mathrm{i}}$ as $\mathbf{W}^{\mathrm{\circ d}}_{\mathrm{i}}$, we focus on the Taylor expansion with respect to variable $\mathbf{W}^{\mathrm{d}}_{\mathrm{i}}$ that
	\begin{equation}\label{eq11}
	\begin{split}
	L&=L(\mathbf{W}^{\mathrm{\circ d}}_{\mathrm{i}}) + 
	(\mathbf{W}^{\mathrm{d}}_{\mathrm{i}} - \mathbf{W}^{\mathrm{\circ d}}_{\mathrm{i}})^{\mathrm{T}} \bigcdot
	\frac{\partial L}{\partial \mathbf{W}^{\mathrm{\circ d}}_{\mathrm{i}}} \ \ \\
	&+\frac{1}{2}(\mathbf{W}^{\mathrm{d}}_{\mathrm{i}} - \mathbf{W}^{\mathrm{\circ d}}_{\mathrm{i}})^{\mathrm{T}}  \bigcdot \frac{\partial^2 L}{\partial (\mathbf{W}^{\mathrm{d}}_{\mathrm{i}})^2} \bigcdot (\mathbf{W}^{\mathrm{d}}_{\mathrm{i}} - \mathbf{W}^{\mathrm{\circ d}}_{\mathrm{i}}) \ \ \\
	&+\mathrm{o}( \ (\mathbf{W}^{\mathrm{d}}_{\mathrm{i}} - \mathbf{W}^{\mathrm{\circ d}}_{\mathrm{i}})^{\mathrm{T}} \bigcdot (\mathbf{W}^{\mathrm{d}}_{\mathrm{i}} - \mathbf{W}^{\mathrm{\circ d}}_{\mathrm{i}}) \ ),
	\end{split} 	
	\end{equation}	
	where $\mathrm{o}( \cdot )$ represents the higher order infinitesimal. Since we fix the range of $\mathbf{W}^{\mathrm{d}}_{\mathrm{i}}$ within $[\alpha, \beta]$, the expectation values of $|\mathbf{W}^{\mathrm{d}}_{\mathrm{i}} - \mathbf{W}^{\mathrm{\circ d}}_{\mathrm{i}}|$ is $(\beta-\alpha)/2$, which is 0.25 in the implementation. Noticing that $\frac{\partial L}{\partial \mathbf{W}^{\mathrm{\circ d}}_{\mathrm{i}}}$ is a null matrix, the second order term in the Taylor expansion would dominate the calculations. Then, with $\frac{\partial^2 L}{\partial (\mathbf{W}^{\mathrm{d}}_{\mathrm{i}})^2}$ being a positive definite matrix, the loss function $L$ approximates a convex function to some extent with respect to $\mathbf{W}^{\mathrm{d}}_{\mathrm{i}}$. The convexity would make it easier to search the global optimums with freely valuing $\mathbf{W}^{\mathrm{d}}_{\mathrm{i}}$ by adjusting $\mathbf{H}_{\hat{\mathrm{d}}}$ if $\mathcal{H}_{fh}$ contains the global optimums.
	
	\textbf{Consequently, HPS-Net could provide more probability to obtain promising results due to the better manifold for parameter optimization, which can make HPS-Net obtain better pixel-wise path selections and better depict pixel-wise land-cover distributions during the inference procedure.}
	However, compared with parameter optimization of conventional deep structures without path selections, Eq.\eqref{eq10} demonstrates that gradient flows of HPS-Net are considerable small due to the extra multiplication of $\mathbf{W}^{\mathrm{d}}_{\mathrm{i}}$, which might make hidden path selection more effective under small training costs compared with larger training costs.
	
	\subsection{Loss Function}\label{Loss_Function}	 
	We utilize fully supervised learning to optimize parameters $\bm{\theta}_f$ in the proposed HPS-Net. Denoting $\mathcal{G}_{\mathbf{I}}$ as the pixel-wise ground truth of $\mathbf{I}$, which indicates the semantic category for each pixel, we have 
	\begin{equation}\label{eq12}
	L = Cross\_Entropy(\mathcal{M}_{\mathbf{I}}, \mathcal{G}_{\mathbf{I}}),
	\end{equation}
	where the cross entropy function takes the model output $\mathcal{M}_{\mathbf{I}}$ in Eq.\eqref{eq8} and ground truth $\mathcal{G}_{\mathbf{I}}$ as the input. Stochastic gradient descent (SGD) is then used in the optimization process of parameter $\bm{\theta}_f$, through which the mini-branch and main-branch can be trained simultaneously.	
	
	\begin{figure*}[!t]
		\centering
		{\includegraphics[width=1\linewidth]{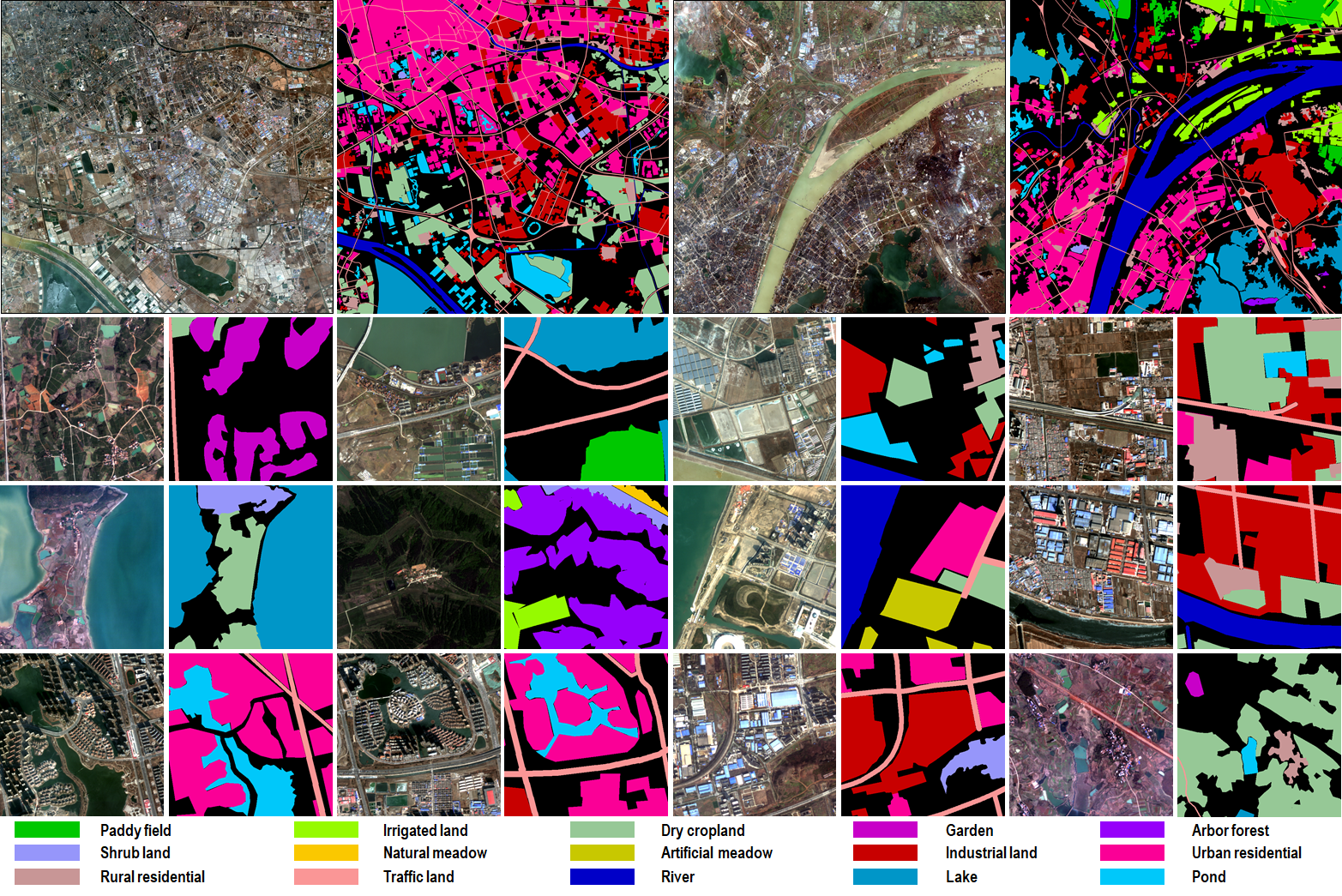}}
		\vspace{-3mm}
		\caption{
			Data samples taken from our proposed GID-15 dataset. Different colors indicate different land-cover categories. The obscure land covers are blackened in the annotation masks. Several intact data samples are listed at the top and some cropped data patches are listed on the bottom, where we can see the annotation meticulousness with wide geographical areas covered. }
		\label{Fig5}
		\vspace{-2mm}
	\end{figure*} 	
	
	\section{The GID-15 Dataset}\label{dataset}
	As the basis for training and evaluation, a well-annotated benchmark dataset is important to develop remote sensing semantic segmentation algorithms.
	Although some widely used datasets with several land-cover categories \cite{2012The, Devis20172017, 20182018, 2017Learning} are provided, remote sensing semantic segmentation datasets are hard to give considerations to both the quantities of data samples and sufficiencies of land-cover categories. As a consequence, the increasing demands in real-world applications, such as identifying {\em industrial land} and {\em urban residential} simultaneously, are difficult to satisfy.
	For instance, containing abundant data samples that cover dozens of cities in China with 150 images of 6800 $\times$ 7200 pixels, GID-5 \cite{DBLP:conf/igarss/TongLXZ18} ensures the data diversity for deep model training,  while the only 5 involved land-cover categories ({\ie} {\em built-up}, {\em farmland}, {\em forest}, {\em meadow} and {\em water}) are not sufficient for real applications. In order to propose a better benchmark meeting practical demands, we expand the fine land-cover classification set in \cite{2020Land} by subdividing 5 land-cover categories in the GID-5 to compose a new GID-15. 
	
	Specifically, to further distinguish sub-categories belonging to the same general land-cover categories, we refer to the Chinese Land Use Classification Criteria (GB/T21010-2017) and finally determine 15 concerned categories through a hierarchical category
	relationships.
	Concretely, for researching the detailed urban area distributions, we subdivide the {\em built-up} into {\em industrial land}, {\em urban residential}, {\em rural residential} and {\em traffic land}. Then, the {\em forest} is subdivided into {\em garden land}, {\em arbor forest} and {\em shrub land} for the researches on vegetation covers, while the {\em meadow} is subdivided into {\em natural meadow} and {\em artificial meadow}. Last but not the least, the {\em farmland} is subdivided into {\em paddy field}, {\em irrigated land} and {\em dry cropland} for researching agricultural land distributions, while the {\em water} is subdivided into {\em river}, {\em lake} and {\em pond} for the water resources researching. 
	
	In practice, an expert group checks the original images and annotated category maps in GID-5 simultaneously, through which we recheck the previous annotations and further separate the annotated land-cover regions in GID-5 according to the hierarchical category relationships we build. Similar with GID-5, all the pixels related to obscure land-cover distributions, which are difficult to depict with pixel-wise land-cover categories explicitly, are ignored in GID-15.
	Further due to the wide range areas covered in GID-5, GID-15 can serve as a challenging benchmark containing large quantities of data samples ({\ie} 150 images of 6800 $\times$ 7200 pixels with pixel-wise annotations) and abundant land-cover categories ({\ie} 15 land-cover categories better meeting practical demands) simultaneously. 
	Several samples of GID-15 are displayed in Fig.\ref{Fig5}, where we can see the annotation meticulousness. The GID-15 dataset would soon be available at \textcolor{blue}{\url{https://captain-whu.github.io/HPS-Net/}}.
	
	\section{Experiments and Analysis}\label{Experiments_and_Analysis}	
	In this section, we evaluate the proposed HPS-Net on both GID-5 and GID-15, where we show the effectiveness of the proposed modules and verify the mathematical analyses in Sec.\ref{Method}. We first illustrate the evaluation metric in Sec.\ref{metric}, while we further clarify the detail experiment settings in Sec.\ref{Exp_Settings}. Then, in Sec.\ref{Basic Architecture Analysis} and Sec.\ref{STOA}, we discuss the application of the proposed structures on different classical deep structures \cite{Resnet} and the state-of-the-art pixel-wise dynamic algorithm \cite{DBLP:journals/tip/GengZQHYZ21}, where we explore the merits of proposed hidden path selection. Further in Sec.\ref{Ablation}, we present detail verifications on the mathematical analyses in Sec.\ref{Method} and study the effects of each term in the proposed structures through ablation studies and feature visualizations.
	
	\subsection{Evaluation Metrics}\label{metric}
	To reduce the influence of label imbalance, we follow the principle in \cite{DBLP:journals/tip/GengZQHYZ21} that utilize mean Intersection over Union (mIoU). Denoting the confusion matrix calculated based on $\mathcal{M}_{\mathbf{I}}$ and $\mathcal{G}_{\mathbf{I}}$ as $\mathrm{Q}=\{\mathrm{q}_{\mathrm{ij}}\}$, we have
	\begin{align}
	\label{eq13}
	\textrm{mIoU} = \frac{1}{\mathrm{N}} \sum\limits^{\mathrm{N}}_{\mathrm{i}=1}{\mathrm{q}_{\mathrm{ii}}/[\sum\limits^{\mathrm{N}}_{\mathrm{j}=1}{(\mathrm{q}_{\mathrm{ij}} + \mathrm{q}_{\mathrm{ji}})}-\mathrm{q}_{\mathrm{ii}}]},
	\end{align}	
	where $\mathrm{N}$ is the total number of land-cover categories. With each term on the right side of Eq.\eqref{eq13} measuring the identification of one certain category, the average form ensures the equivalence of each category.
	
	\subsection{Experiment Settings}\label{Exp_Settings}
	The deep structures and semantic segmentation algorithms involved in our experiments are as follows:
	\begin{itemize}
		\item[-] Resnet-50~\cite{Resnet}: a deep residual structure serving as the main branch in HPS-Net, where 16 skip connections are considered to be applied hidden path selection. 
		
		\item[-] Resnet-101~\cite{Resnet}: a deeper version of residual structure compared with Resnet-50, where 33 skip connections are considered to be applied hidden path selection. 
		
		\item[-] Pspnet~\cite{PSPnet}: a semantic segmentation algorithm that utilizes multi-scale structures based on poolings of different size.
		
		\item[-] Deeplabv3*~\cite{V3+}: a semantic segmentation algorithm taking residual structures as the backbone, which integrates the encoder-decoder structures and multi-scale structures based on dilation convolutions.
		
		\item[-] GPSNet~\cite{DBLP:journals/tip/GengZQHYZ21}: a semantic segmentation algorithm that dynamically selects adaptive forward paths for each pixel.
		\color{black}
	\end{itemize}
	
	In this paper, we choose residual structures as the main-branch in HPS-Net. All the experiments are implemented on a Pascal V100 with 16G memory based on Pytorch. All the model training and evaluation are under the same conditions without pre-trained parameters and post-processings.
	In the training process, SGD is utilized with batch size as 10 and random flip is utilized as the data augmentation. The initial learning rate is set as 0.007 and poly policy is employed with the power of 0.9. We set the momentum as 0.9 and weight decay as 0.0001. In the testing process, we take raw outputs of each model as the results for the evaluation.
	
	As for the hyper-parameters in HPS-Net, we set $\mathscr{L}_{\mathrm{d}}=\{\mathrm{d}_\mathrm{k}\}_{k \in \mathbb{N}_+}$ as the index of layers containing skip connections, which indicates that total 16 layers in Resnet-50 and 33 layers in Resnet-101 are applied hidden path selection in the main-branch. Then, with residual structures divided into several stages according to the pixel resolution, we set $\mathscr{H}_{\mathrm{d}}=\{\mathrm{d}_\mathrm{m}\}_{m \in \mathbb{N}_+}$ as the index of last layer in each stage, which indicates that total 4 layers serve as the hidden variables derived from the mini-branch.
	Finally, we generally set $(\alpha, \beta)$ as (0.75, 1.25) in the HP-Module, where the softmax function is multiplied with 2 to avoid the gradient vanishing. Especially, in the first layer of each stage in the main-branch, $(\alpha, \beta)$ are expanded to (0.5, 1.5) due to the additional alternative paths from previous stages.
	
	For all the experiments in both GID-5 and GID-15, we crop the images and corresponding annotation masks into patches of 512 $\times$ 512 pixels simultaneously, which are splitted into two subsets, {\ie} totally 25200 data patches for training and 6300 data patches for testing.
	
	\begin{figure}[!t]
		\centering
		{\includegraphics[width=1.0\linewidth]{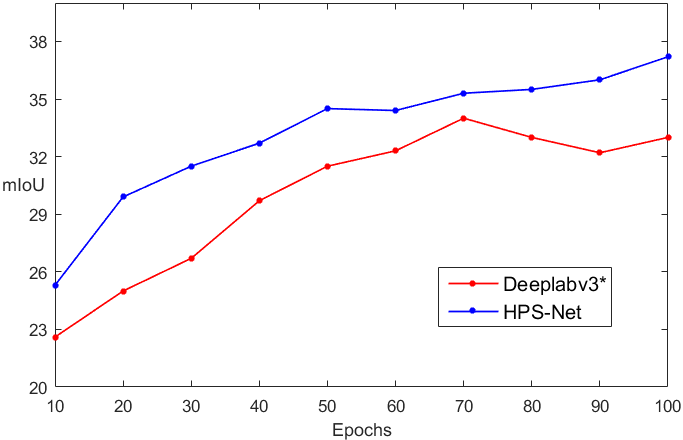}}
		\vspace{-3mm}
		\caption{Toy experiments involving in randomly selected 10$\%$ data samples. We train the Deeplabv3* and HPS-Net, where hidden path selection is applied on Resnet-101 in Deeplabv3*, for different training epochs. The larger mIoU of HPS-Net indicates the merits of hidden path selection and the increasing trendency of fold line corresponding to HPS-Net implies the better shape of high dimension manifold, which makes global optimums more accessible.
		} 
		\label{Fig6}
		\vspace{-2mm}
	\end{figure} 
	
	\begin{figure*}[!t]
		\centering
		{\includegraphics[width=1\linewidth]{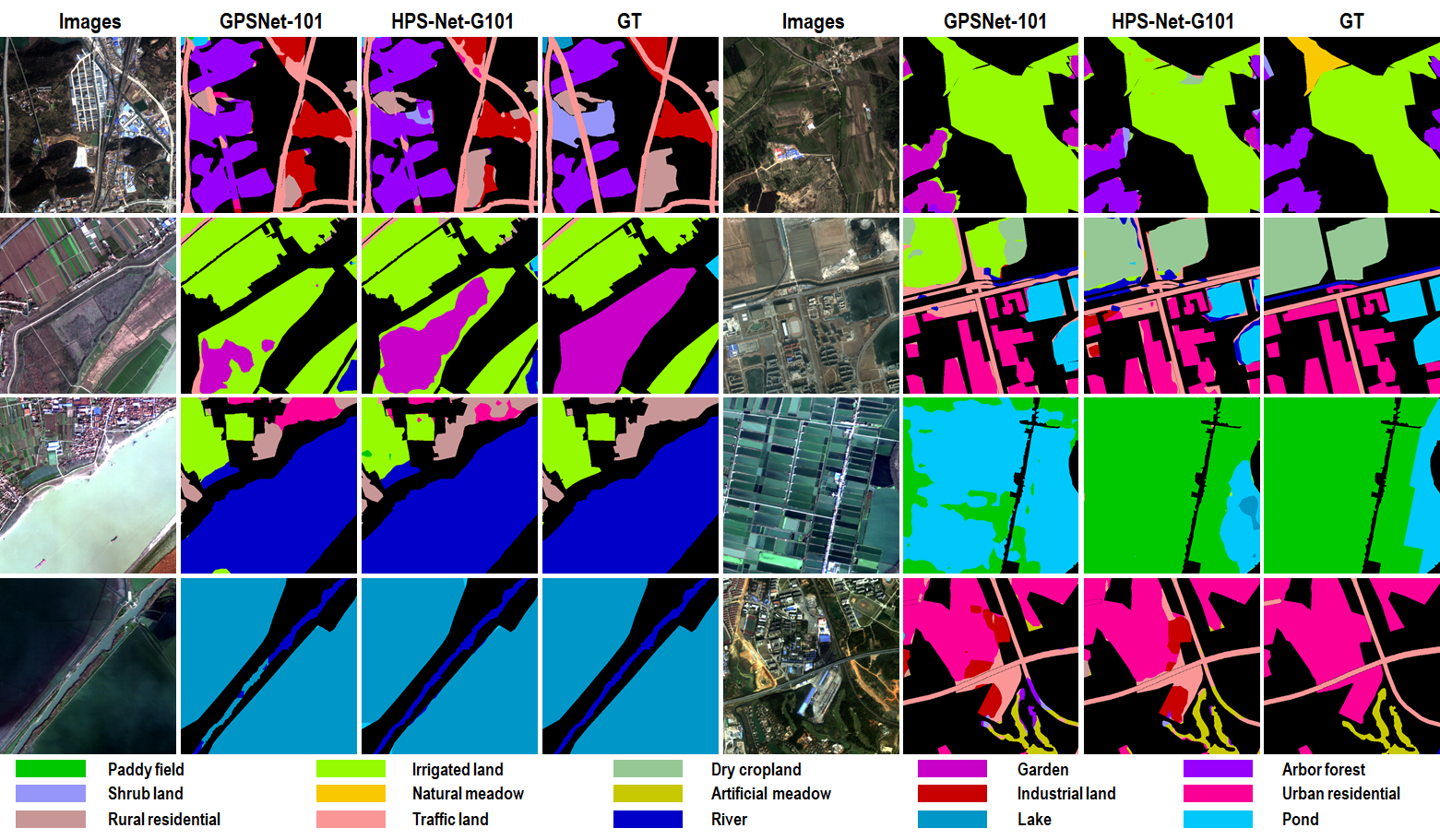}}
		\vspace{-3mm}
		\caption{
			Examples of comparing our HPS-Net with \textcolor{black}{the} state-of-the-art pixel-wise dynamic method GPSNet on GID-15 with Resnet-101 as the backbone.
			Our proposed HPS-Net can better depict pixel-wise land-cover distributions. The blackened pixels are the ignored regions in the annotation masks, which are also ignored in the semantic segmentation results.}
		\label{Fig7}
		\vspace{-2mm}
	\end{figure*} 
	
	\begin{table*}[htb!]
		\scriptsize
		\centering
		\renewcommand\tabcolsep{3.9pt} 
		\small
		\caption{Comparisons with state-of-the-art methods trained on GID-15. We denote applying the hidden path selection on Deeplabv3* as HPS-Net-101 and HPS-Net-50, while applying hidden path selection on GPSNet with different backbones are denoted as HPS-Net-G101 and HPS-Net-G50. Number 50 and number 101 following the model name indicate the Resnet-50 and Resnet-101 backbones. Land-cover categories listed in the first row are abbreviations following the order in Fig.\ref{Fig5}.}
		\vspace{-3mm}
		\label{tab:GID-15}
		\setlength{\arraycolsep}{0.1pt}
		\begin{threeparttable}
			\begin{tabular}{c|ccccccccccccccc|c}
				\toprule
				
				\multirow{4}{*}{{Methods}} &\multirow{4}{*}{\rotatebox{70}{P. field}}&\multirow{4}{*}{\rotatebox{67}{Irr. land}}&\multirow{4}{*}{\rotatebox{67}{Dry cropl.}}&\multirow{4}{*}{\rotatebox{67}{Garden}}&\multirow{4}{*}{\rotatebox{67}{Arb. forest}}&\multirow{4}{*}{\rotatebox{67}{Shr. land}}&\multirow{4}{*}{\rotatebox{67}{Nat. mead.}}&\multirow{4}{*}{\rotatebox{67}{Art. mead.}}&\multirow{4}{*}{\rotatebox{67}{Ind. land}}&\multirow{4}{*}{\rotatebox{67}{Urb. resid.}}&\multirow{4}{*}{\rotatebox{67}{Rur. resid.}}&\multirow{4}{*}{\rotatebox{67}{Traff. land}}&\multirow{4}{*}{\rotatebox{67}{River}}&\multirow{4}{*}{\rotatebox{67}{Lake}}&\multirow{4}{*}{\rotatebox{67}{Pond}}&\multirow{4}{*}{{mIoU}} \\
				
				& & & & & & & & & & & & & & & & \\
				
				& & & & & & & & & & & & & & & & \\
				
				& & & & & & & & & & & & & & & & \\
				
				\midrule
				
				PSPNet~\cite{PSPnet}&54.4&74.3&40.1&21.4&85.6&11.9&68.4&30.4&62.9&73.4&61.4&54.9&57.5&68.0&\textbf{28.7}&52.9\\
				Deeplabv3*-50~\cite{V3+}&\textbf{56.6}&70.2&41.1&21.2&89.2&15.6&64.5&37.9&64.1&73.9&62.5&59.3&\textbf{60.9}&\textbf{73.5}&23.6&54.3\\
				
				\color{black} Deeplabv3*-101~\cite{V3+}&\textcolor{black}{55.7}&\textcolor{black}{74.2}&\textcolor{black}{44.0}&\textcolor{black}{19.5}&\textcolor{black}{89.3}&\textcolor{black}{12.2}&64.0&38.9&64.4&74.4&63.7&57.6&55.2&70.7&27.9&54.1\\
				\color{black}
				GPSNet-50~\cite{DBLP:journals/tip/GengZQHYZ21}&54.6&72.3&41.1&23.1&90.1&15.6&66.5&38.8&63.6&74.1&63.9&57.1&55.6&71.9&26.7&54.3\\
				GPSNet-101~\cite{DBLP:journals/tip/GengZQHYZ21}&53.4&73.4&43.8&16.8&88.8&16.5&66.2&42.0&63.2&73.4&61.2&56.7&52.9&71.5&25.6&53.7\\		
				\hline
				HPS-Net-50 (Ours) &54.1&73.4&41.9&\textbf{24.3}&90.1&13.2&66.7&40.7&\textbf{65.2}&\textbf{75.0}&\textbf{65.2}&\textbf{60.8}&57.1&71.4&26.4&55.0\\
				HPS-Net-101 (Ours) &55.2&{74.7}&\textbf{47.6}&{20.8}&{88.7}&{12.7}&63.9&41.3&64.7&74.6&63.1&59.5&58.9&71.5&25.8&54.9\\
				HPS-Net-G50 (Ours) &56.1&74.0&47.0&20.4&91.0&\textbf{17.3}&68.8&44.0&63.9&74.7&\textbf{65.2}&59.7&52.1&71.5&26.5&\textbf{55.5}\\
				HPS-Net-G101 (Ours) &{53.2}&\textbf{75.6}&{47.3}&{22.1}&\textbf{91.1}&{17.0}&\textbf{69.5}&\textbf{45.2}&63.3&74.0&64.4&59.0&55.6&72.4&23.4&\textbf{55.5}\\	
				\bottomrule
			\end{tabular}
		\end{threeparttable}
	\end{table*}
	
	\subsection{Discussions on Hidden Path Selection}\label{Basic Architecture Analysis}	
	For simply discussing the application of hidden path selection on existing deep structures, we first run a toy experiment, where 10$\%$ data samples are randomly selected from GID-15 for training and evaluation. Specifically, we apply the hidden path selection on Resnet-101 in Deeplabv3* to compose HPS-Net, which is utilized to compare with the original residual structure. The experimental results with different epochs, {\ie} from 10 epochs to 100 epochs at 10 intervals, are shown in Fig.\ref{Fig6}, where we can see the fold lines marking evaluated mIoU corresponding to each model with the same Resnet-101 backbone.
	
	
	Specifically, as for the fold line corresponding to Deeplabv3*, we can see that the model encounters a bottleneck around 32.0 in mIoU. However, with only 10 $\%$ data samples selected, the model training is far from saturation. Thus, this bottleneck indicates that the parameter optimization in Deeplabv3* on the high dimension manifold hovers at the same level, which implies the defects of aforementioned manifold. Then, for the fold line corresponding to HPS-Net, the model performance is constantly improved with increasing training epochs, which implies the better shape of manifold for parameter optimization in HPS-Net to make global optimums more accessible as demonstrated by the mathematical analyses in Sec.\ref{Model_Analyses}. 
	As for the comparison of two fold lines, HPS-Net outperforms Deeplabv3* by average 3.2 in mIoU, which indicates the improvement attributed to applying hidden path selection on the basic residual structure.
	
	\subsection{Comparison with the State-of-the-art}\label{STOA}	
	To compare with state-of-the-art algorithms, we list the IoU for each land-cover category and mIoU for overall evaluation in Tab.\ref{tab:GID-15}, where all models are trained on GID-15 with 20 epochs from scratch. 
	We apply the hidden path selection on Deeplabv3* and GPSNet to compose the HPS-Net. We denote applying the hidden path selection on Deeplabv3* as HPS-Net-50 and HPS-Net-101, while applying hidden path selection on GPSNet with different backbones are denoted as HPS-Net-G50 and HPS-Net-G101. Number 50 and number 101 following the model name indicate the Resnet-50 and Resnet-101 backbones.
	As we can see, HPS-Net tends to achieve better results on the majority of land-cover categories, while the categories related to {\em water} can not be depicted that well. Furthermore we can conclude from Tab.\ref{tab:GID-15} that HPS-Net-G50 and HPS-Net-G101 achieve the best results, {\ie} 55.5 in mIoU.
	
	\begin{figure*}[!t]
		\centering
		{\includegraphics[width=1\linewidth]{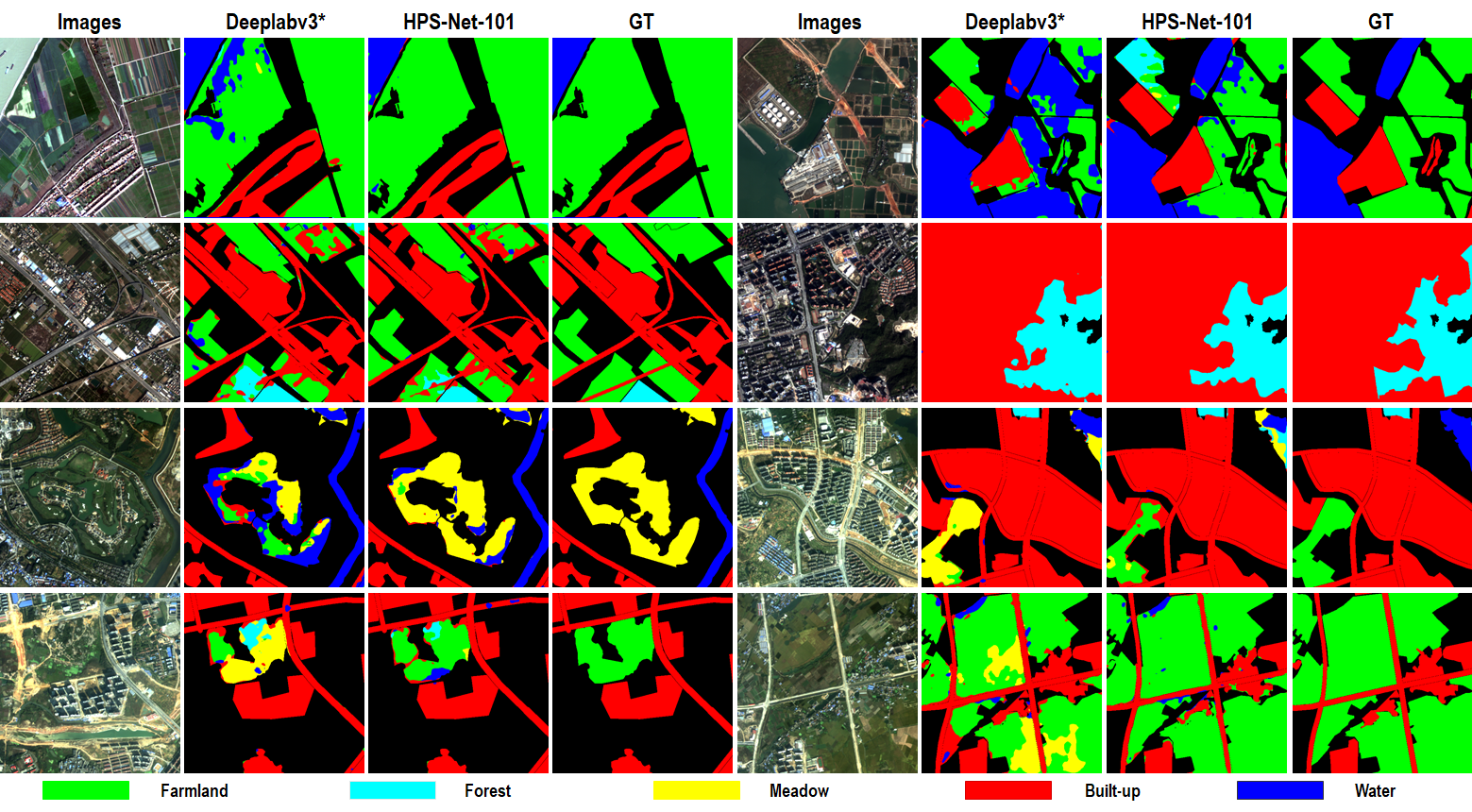}}
		\vspace{-3mm}
		\caption{
			Examples of comparing our HPS-Net with the basic residual structures Deeplabv3* on GID-5 with Resnet-101 as the backbone, which display the merits of applying the proposed hidden path selection on Deeplabv3* ({\ie} HPS-Net-101).}
		\label{Fig8}
		\vspace{-2mm}
	\end{figure*} 	
	
	\begin{table}[htb!]
		\scriptsize
		\centering
		\renewcommand\tabcolsep{3.9pt} 
		\small
		\caption{Comparisons with state-of-the-art methods on GID-5.}
		\vspace{-3mm}
		\label{tab:GID-5}
		\setlength{\arraycolsep}{0.1pt}
		\begin{threeparttable}
			\begin{tabular}{c|ccccc|c}
				\toprule
				
				\multirow{4}{*}{{Methods}} &\multirow{4}{*}{\rotatebox{69}{Farmland}}&\multirow{4}{*}{\rotatebox{67}{Forest}}&\multirow{4}{*}{\rotatebox{67}{Meadow}}&\multirow{4}{*}{\rotatebox{67}{Built-up}}&\multirow{4}{*}{\rotatebox{67}{Water}}&\multirow{4}{*}{{mIoU}} \\
				
				& & & & & &\\
				
				& & & & & &\\
				
				& & & & & & \\
				
				\midrule
				
				PSPNet~\cite{PSPnet}&89.7&87.2&53.7&93.8&\textbf{92.1}&83.3\\
				Deeplabv3*-50~\cite{V3+}&\textbf{90.1}&87.5&53.2&94.1&91.8&83.3\\
				
				\color{black} Deeplabv3*-101~\cite{V3+}&89.4&86.9&54.3&93.7&91.1&83.1\\
				\color{black}
				GPSNet-50~\cite{DBLP:journals/tip/GengZQHYZ21}&90.0&88.0&55.6&93.7&92.0&83.9\\
				GPSNet-101~\cite{DBLP:journals/tip/GengZQHYZ21}&88.6&86.7&51.9&91.6&91.0&82.0\\		
				\hline
				HPS-Net-50 (Ours) &90.0&86.7&57.0&94.4&91.2&83.8\\
				HPS-Net-101 (Ours) &\textbf{90.1}&87.6&57.1&\textbf{94.5}&92.0&\textbf{84.3}\\
				HPS-Net-G50 (Ours) &\textbf{90.1}&87.0&\textbf{57.2}&93.9&91.4&83.9\\
				HPS-Net-G101 (Ours) &90.0&\textbf{88.2}&55.8&94.0&91.9&84.0\\ 
				
				\bottomrule
			\end{tabular}
		\end{threeparttable}
	\end{table}
	
	Specifically, HPS-Net-50 outperforms Deeplabv3*-50 by 0.7 in mIoU and HPS-Net-101 raises 0.8 improvement over Deeplabv3*-101, while HPS-Net-G50 outperforms GPSNet-50 by 1.2 in mIoU and HPS-Net-G101 raises 1.8 improvement over GPSNet-101. The model comparisons listed above illustrate that the applied hidden path selection can stably improve the performance of existing deep structures.
	Besides, we can check that GPSNet-50 and GPSNet-101 achieve comparative results with Deeplabv3*-50 and Deeplabv3*-101, which indicates that the proposed hidden path selection can build better pixel-wise dynamic structures than the gate path selection proposed in the state-of-the-art algorithm GPSNet. However, together with the application of hidden path selection, gate path selection is able to raise more stable improvements according to the comparison between HPS-Net-50 and HPS-Net-G50 or HPS-Net-101 and HPS-Net-G101.
	We display the semantic segmentation results in Fig.\ref{Fig7}, where we can also see the superiorities of the proposed HPS-Net. 
	
	For the experiments on GID-5, we recover the 5 land-cover categories in GID-5 from the data samples involved in experiments mentioned above according to the hierarchical category relationships in Sec.\ref{dataset}, based on which all models are also trained for 20 epochs from scratch. The experimental results are summarized in Tab.\ref{tab:GID-5}, where we can also see that HPS-Net obtains better results. However, it is worth noticing that, the gaps between HPS-Net and other algorithms are smaller than that in the experiment results on GID-15, which demonstrates that GID-15 can better evaluate different semantic segmentation algorithms. We also illustrate the semantic segmentation results in Fig.\ref{Fig8} to show the merits of applying hidden path selection on the basic residual structures.
	
	
	\begin{figure*}[!t]
		\centering
		{\includegraphics[width=1\linewidth]{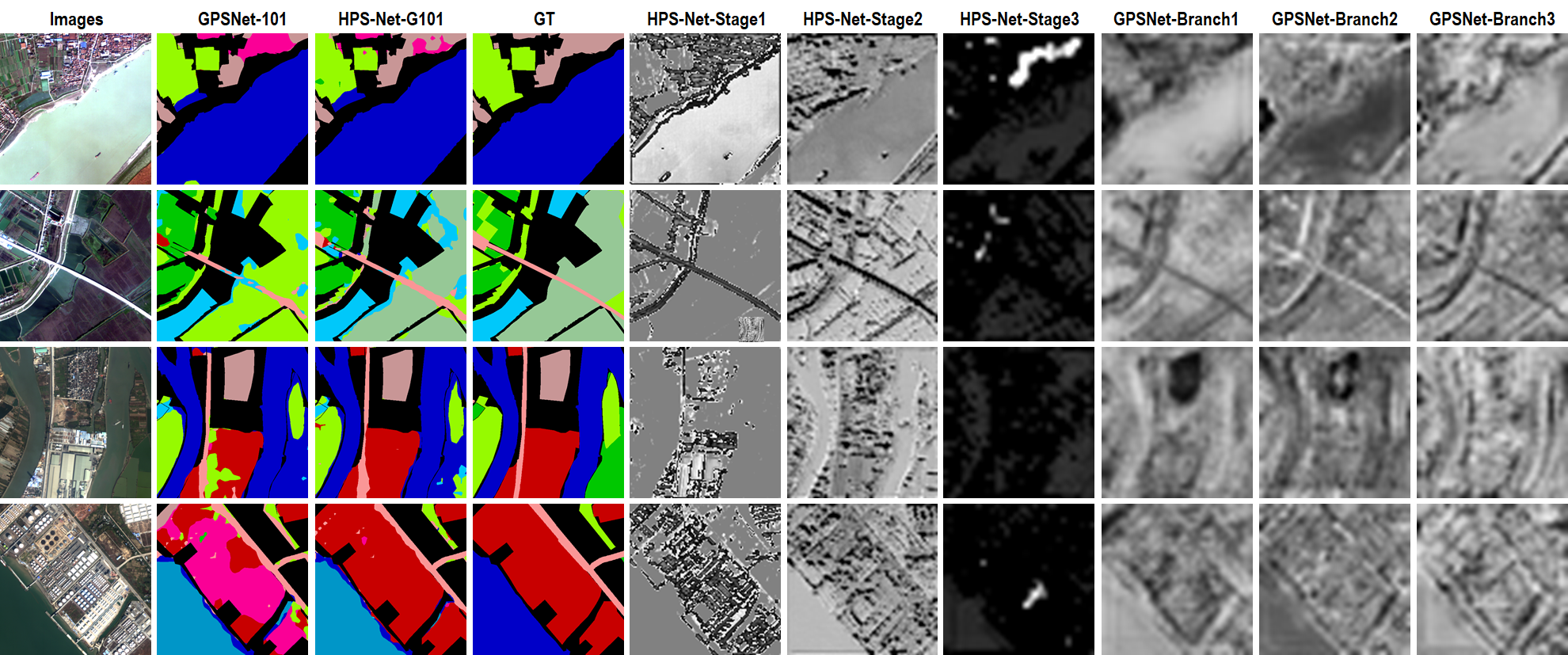}}
		\vspace{-3mm}
		\caption{
			Visualization of the masks corresponding to path selection in HPS-Net-G101 and GPSNet-101. We visualize the path selection for the first layer of 1-st stage to 3-rd stage in HPS-Net-G101, while the first mask of 1-st branch to 3-rd branch in GPSNet are listed too. The brighter grayscale values in these masks indicate larger factors corresponding to the first alternative path, where the proposed HPS-Net can better depict pixel-wise land-cover distributions through diverse forward path selections for different pixels compared with GPSNet.}
		\label{Fig9}
		\vspace{-2mm}
	\end{figure*} 
	
	\begin{table*}[htb!]
		\scriptsize
		\centering
		\renewcommand\tabcolsep{3.9pt} 
		\small
		\caption{Ablation studies on GID15. HPS-Net-50-ps and HPS-Net-101-ps represent all pixels sharing the same dynamic factor. HPS-Net-50-ig and HPS-Net-101-ig represent intact gradient flows in HP-Module. HPS-Net-50-fh and HPS-Net-101-fh represent fixing the hidden variables to null tensors in HP-Module.}
		\vspace{-3mm}
		\label{tab:ablation}
		\setlength{\arraycolsep}{0.1pt}
		\begin{threeparttable}
			\begin{tabular}{c|ccccccccccccccc|c}
				\toprule
				
				\multirow{4}{*}{{Methods}} &\multirow{4}{*}{\rotatebox{70}{P. field}}&\multirow{4}{*}{\rotatebox{67}{Irr. land}}&\multirow{4}{*}{\rotatebox{67}{Dry cropl.}}&\multirow{4}{*}{\rotatebox{67}{Garden}}&\multirow{4}{*}{\rotatebox{67}{Arb. forest}}&\multirow{4}{*}{\rotatebox{67}{Shr. land}}&\multirow{4}{*}{\rotatebox{67}{Nat. mead.}}&\multirow{4}{*}{\rotatebox{67}{Art. mead.}}&\multirow{4}{*}{\rotatebox{67}{Ind. land}}&\multirow{4}{*}{\rotatebox{67}{Urb. resid.}}&\multirow{4}{*}{\rotatebox{67}{Rur. resid.}}&\multirow{4}{*}{\rotatebox{67}{Traff. land}}&\multirow{4}{*}{\rotatebox{67}{River}}&\multirow{4}{*}{\rotatebox{67}{Lake}}&\multirow{4}{*}{\rotatebox{67}{Pond}}&\multirow{4}{*}{{mIoU}} \\
				
				& & & & & & & & & & & & & & & & \\
				
				& & & & & & & & & & & & & & & & \\
				
				& & & & & & & & & & & & & & & & \\
				
				\midrule
				
				HPS-Net-50-ps&52.5&72.9&42.0&  {24.4}&88.2&  {15.3}&64.1&40.7&64.8&74.9&64.5&60.2&  {59.5}&69.6&24.3&54.5\\
				HPS-Net-101-ps&54.4&72.8&46.3&19.4&89.9&14.4&61.5&33.6&59.4&73.6&61.3&55.4&50.7&68.2&  {29.2}&52.7 \\
				\hline
				HPS-Net-50-fh&52.0&71.8&37.9&22.8&85.7&12.8&63.1&39.2&63.9&74.5&63.8&59.9&53.8&65.5&22.3&52.6\\
				HPS-Net-101-fh&57.1&72.4&40.6&20.6&87.5&12.5&62.9&34.7&62.8&73.6&63.5&57.6&48.9&64.6&22.7&52.1\\
				\hline
				HPS-Net-50-ig&  {56.0}&71.2&39.9&21.4&86.2&13.0&65.0&38.9&63.6&74.0&63.6&58.4&57.4&70.3&25.2&53.6\\
				HPS-Net-101-ig&55.9&72.2&38.0&20.6&88.9&14.5&62.3&  {41.8}&64.5&74.1&63.0&58.5&47.9&68.3&25.7&53.1\\
				\hline
				HPS-Net-50 &54.1&73.4&41.9&24.3&  {90.1}&13.2&  {66.7}&40.7&  {65.2}&  {75.0}&  {65.2}&  {60.8}&57.1&71.4&26.4&  {55.0}\\
				HPS-Net-101 &55.2&  {74.7}&  {47.6}&20.8&88.7&12.7&63.9&41.3&64.7&74.6&63.1&59.5&58.9&  {71.5}&25.8&54.9\\
				\bottomrule
			\end{tabular}
		\end{threeparttable}
	\end{table*}
	
	\subsection{Ablation Study and Feature Visualization}\label{Ablation}	
	To study the effects of each term in the proposed structures, we design several ablation studies in this section. Firstly, we average the mask $\mathbf{W}^{\mathrm{d}}_{\mathrm{i}}$ in HPS-Net and make all pixels sharing the same path selection, denoted as HPS-Net-50-ps and HPS-Net-101-ps. Subsequently, we can explore the merits of pixel-wise dynamic structures compared with image-wise dynamic structures. Then, we fix the hidden variables in HPS-Net, denoted as HPS-Net-50-fh and HPS-Net-101-fh, to study the influence of direct relationships between $\mathbf{W}^{\mathrm{d}}_{\mathrm{i}}$ and $\mathbf{F}_{\mathrm{i}}$ analyzed in Sec.\ref{Gated_Path_Selection}. Besides, we further keep intact gradient flows in the HP-Module, denoted as HPS-Net-50-ig and HPS-Net-101-ig, to study the gradient analyses in Sec.\ref{Hidden_Selection_Network}. 
	
	As illustrated in Tab.\ref{tab:ablation}, HPS-Net-50-ps and HPS-Net-101-ps are inferior to HPS-Net-50 and HPS-Net-101 by 0.5 and 2.2 in mIoU respectively, which indicates the superiority of pixel-wise dynamic structures.
	Then, HPS-Net-50 and HPS-Net-101 outperform HPS-Net-50-fh and HPS-Net-101-fh by 2.4 and 2.8 in mIoU respectively, which verifies the deleteriousness of direct relationships between $\mathbf{W}^{\mathrm{d}}_{\mathrm{i}}$ and $\mathbf{F}_{\mathrm{i}}$ analyzed in Sec.\ref{Gated_Path_Selection}.	Finally, HPS-Net-50-ig and HPS-Net-101-ig achieve lower results by 1.4 and 1.8 in mIoU than HPS-Net-50 and HPS-Net-101 respectively, which verifies the inefficiency of model training with intact gradient flows analyzed in Sec.\ref{Hidden_Selection_Network}.

	For illustrating the depiction of pixel-wise land-cover distributions by HPS-Net during the inference procedure, we visualize the selected paths of several data samples in Fig.\ref{Fig9}, where the proposed HPS-Net can better depict pixel-wise land-cover distributions through dynamic pixel-wise forward paths compared with GPSNet. Specifically, as illustrated in the first row of Fig.\ref{Fig9}, HPS-Net emphasizes paths with more complex computations, {\ie} the first alternative path, among rural residential at top of the image, which alleviates the mis-prediction compared with GPSNet. Moreover, as shown in the second row, the combination of path selections from 1-st stage to 3-rd stage ensures different forward paths for irrigated land at left-top and dry cropland at right-bottom, which makes efforts to distinguish these two land-cover categories in the same image and also alleviate the mis-prediction. As we can see, the proposed HPS-Net can better depict pixel-wise land-cover distributions during the inference procedure.
	
	\begin{table}[htb!]
		\scriptsize
		\centering
		\renewcommand\tabcolsep{3.9pt} 
		\small
		\caption{Ablation studies on the computational efficiency of hidden path selection. HPS-Net-w/o-h means removing hidden variables. All the numbers represent FLOPs (G) of each model.}
		\vspace{-3mm}
		\label{tab:Ablation_Efficiency}
		\setlength{\arraycolsep}{0.1pt}
		\begin{threeparttable}
			\begin{tabular}{c|c|c|c}
				\toprule
				
				\multirow{1}{*}{{Backbones}} &\multirow{1}{*}{{Deeplabv3*}}&			\multirow{1}{*}{{HPS-Net-w/o-h}} &\multirow{1}{*}{{HPS-Net}} \\
				
				\midrule

				Resnet-50 &69.2& 74.9&75.2\\
				
				\hline
				Resnet-101 &88.6& 94.6&95.0\\

				\bottomrule
			\end{tabular}
		\end{threeparttable}
	\end{table}			
	
	Finally, to explore the efficiency of proposed HPS-Net, we list the computation costs of each model in Tab.\ref{tab:Ablation_Efficiency}. As we can see, pixel-wise dynamic structures cause 5.7G extra FLOPs and hidden variables result in 0.3G extra FLOPs with Resnet-50 as the backbone, while the corresponding extra FLOPs are 6.0G and 0.4G respectively with Resnet-101 as the backbone. Compared with the FLOPs of Deeplabv3*, HPS-Net results in less than 10$\%$ extra computation costs, which indicates that HPS-Net gives the consideration to both effectiveness and efficiency.

	\section{Conclusion}
	In this paper, we propose a hidden path selection network for semantic segmentation in remote sensing images to select adaptive forward paths for every pixel guided by the mathematical analyses in terms of the parameter optimization. The inherent problem about inaccessible global optimums is tackled with the help of hidden variables derived from an extra mini-branch in HPS-Net. For the better training and evaluation, we expand the land-cover classification dataset, {{\ie} GID-5}, into 15 land-cover categories and propose the new GID-15 dataset, which provides more abundant land-cover categories and better satisfies the real-world scenarios.
	The experimental results on both GID-5 and GID-15 show that the proposed module can stably improve the performances of existing deep structures.

	\ifCLASSOPTIONcaptionsoff
	\newpage
	\fi

	
	
	%
	{\small
		\bibliographystyle{IEEEtran}
		\bibliography{egbib}
	}
	
	%
	
	\ifCLASSOPTIONcaptionsoff
	\newpage
	\vskip -1.5\baselineskip plus -1fil
	
	\begin{IEEEbiography}[{\includegraphics[width=1in,height=1.25in,clip,keepaspectratio]{images/ykp.png}}]{Kunping Yang} received his B.S. degree and M.S. degree in Mathematics from Wuhan University, Wuhan, China, in 2014 and 2016 respectively. He is currently pursuing his Ph.D. degree in the State Key Laboratory of Information Engineering in Surveying, Mapping and Remote Sensing (LIESMARS) at Wuhan University. His research interests include mathematical modeling of images, remote sensing image understanding, semantic segmentation and change detection.
	\end{IEEEbiography}
	
	\vskip -1.9\baselineskip plus -1fil
	
	\begin{IEEEbiography}[{\includegraphics[width=1in,height=1.25in,clip,keepaspectratio]{images/xgs.jpg}}]{Gui-Song Xia}
		(M'10-SM'15) received his Ph.D. degree in image processing and computer vision from CNRS LTCI, T{\'e}l{\'e}com ParisTech, Paris, France, in 2011. From 2011 to 2012, he has been a Post-Doctoral Researcher with the Centre de Recherche en Math{\'e}matiques de la Decision, CNRS, Paris-Dauphine University, Paris, for one and a half years. He has also been working as Visiting Scholar at DMA, {\'E}cole Normale Sup{\'e}rieure (ENS-Paris) in 2018.
		
		He is currently working as a full professor in computer vision and photogrammetry jointly in the School of Computer Science and the State Key Lab. of LIESMARS at Wuhan University, where he is leading a research team named Computational and Photogrammetric Vision Team (CAPTAIN), working toward developing mathematical and computational models to measure and understand our physic environments with vision information. His current research interests include mathematical modeling of images and videos, structure from motion, perceptual grouping, and remote sensing image interpretation. He serves on the  Editorial Boards of the journals Pattern Recognition, Signal Processing: Image Communications, EURASIP Journal on Image \& Video Processing, and Journal of Remote Sensing. He has also served as Guest Editors for journals including IEEE Trans. on Big Data, Pattern Recognition Letter, {\em etc.}.
	\end{IEEEbiography}
	
	\vskip -1.9\baselineskip plus -1fil
	
	\begin{IEEEbiography}[{\includegraphics[width=1in,height=1.25in,clip,keepaspectratio]{images/lzc.png}}]{Zicheng Liu} received his B.S. degree in Electrical Engineering from Wuhan University of Science and Technology, Wuhan, China, in 2018. He is currently pursuing his M.S. degree in the State Key Laboratory of Information Engineering in Surveying, Mapping and Remote Sensing (LIESMARS) at Wuhan University. His research interests include remote sensing image understanding and change detection.
	\end{IEEEbiography}
	
	\vskip -1.9\baselineskip plus -1fil
	
	\begin{IEEEbiography}[{\includegraphics[width=1in,height=1.25in,clip,keepaspectratio]{images/dubo.jpg}}]{Bo Du}
		(M’10–SM’15) received the Ph.D. degree  from the State Key Lab. LIESMARS, Wuhan University, Wuhan, China, in 2010.
		He is currently a professor with the National Engineering Research Center for Multimedia Software, Institute of Artificial Intelligence and School of Computer Science, Wuhan University, China. He has more than 60 research papers published in the IEEE Trans. on Neural Networks and Learning System (TNNLS), IEEE Trans. on image processing (TIP), and IEEE Trans. on Multimedia (TMM), IEEE Trans. on Geoscience and Remote Sensing (TGRS), IEEE Journal of Selected Topics in Earth Observations and Applied Remote Sensing (JSTARS), and IEEE Geoscience and Remote Sensing Letters (GRSL), etc. Thirteen of them are ESI hot papers or highly cited ones. His major research interests include pattern recognition, hyperspectral image processing, machine learning and signal processing.
		
		He is currently a senior member of IEEE. He received 2020 Best Paper award for IEEE TGRS, Highly Cited Researcher 2019 award from Web of Science Group, the distinguished paper award from IJCAI 2018, the best paper award of IEEE Whispers 2018 and the champion award of the IEEE Data Fusion Contest 2018. He received ACM Rising Star Awards Wuhan in 2015. He serves as associate editor for Pattern Recognition and Neurocomputing, senior PC/PC for IJCAI/AAAI/KDD, and as area chair for ICPR and IJCNN. He was the Session Chair for both International Geoscience and Remote Sensing Symposium (IGARSS) 2018/2016 and the 4th IEEE GRSS Workshop on Hyperspectral Image and Signal Processing: Evolution in Remote Sensing. 
	\end{IEEEbiography}
	
	\vskip -1.9\baselineskip plus -1fil
	
	\begin{IEEEbiography}[{\includegraphics[width=1in,height=1.25in,clip,keepaspectratio]{images/wenyang.jpg}}]{Wen Yang}
		(M'09-SM'16) received a B.S. degree in electronic apparatus and surveying technology, a M.S. degree in computer application technology and a Ph.D. degree in communication and information system from Wuhan University, Wuhan, China, in 1998, 2001, and 2004, respectively. From 2008 to 2009, he worked as a Visiting Scholar with the Apprentissage et Interfaces (AI) Team, Laboratoire Jean Kuntzmann (LJK), Grenoble, France. From 2010 to 2013, he worked as a Post-doctoral Researcher with the State Key Lab. LIESMARS, Wuhan University. Since then, he has been a Full Professor with the School of Electronic Information, Wuhan University. His research interests include object detection and recognition, semantic segmentation and multisensor information fusion.
	\end{IEEEbiography}
	
	\vskip -1.9\baselineskip plus -1fil
	
	\begin{IEEEbiography}[{\includegraphics[width=1in,height=1.25in,clip,keepaspectratio]{images/Marcello.png}}]{Marcello Pelillo} (SM'04-F'13) 
		is a Full Professor of Computer Science at Ca’ Foscari University,
		Venice, where he leads the Computer Vision and Pattern Recognition Lab. He has been the Director of the European Centre for Living
		Technology (ECLT) and has held visiting research/teaching positions in several
		institutions including Yale University (USA), University College London (UK), McGill
		University (Canada), University of Vienna (Austria), York University (UK), NICTA
		(Australia), Wuhan University (China), Huazhong University of Science and Technology
		(China), and South China University of Technology (China). He is also
		an external affiliate of the Computer Science Department at Drexel University (USA). 
		His research interests are in the areas of computer vision, machine learning and pattern recognition where he has published more than 200 technical papers in refereed journals, handbooks, and conference proceedings.
		
		He has been General Chair for ICCV 2017, Program Chair for ICPR 2020, and has been Track or Area Chair for several conferences in his area. 
		He is the Specialty Chief Editor of Frontiers in Computer Vision and serves, or has served, on the Editorial Boards of several journals, including IEEE Transactions on Pattern Analysis and Machine Intelligence, Pattern Recognition, IET Computer Vision, and Brain Informatics.
		He also serves on the Advisory Board of Springer’s International Journal of Machine Learning and Cybernetics.
		Prof. Pelillo has been elected Fellow of the IEEE and Fellow of the IAPR and is an IEEE SMC Distinguished Lecturer. His Erd\"os number is 2.
		
	\end{IEEEbiography}
	
	
	
	
	\fi
	
\end{document}